\documentclass[a4paper,12pt,dvipsnames,usernames]{article}
\pdfoutput=1

\usepackage{aas_macros}

\usepackage{jcappub} 

\usepackage[utf8]{inputenc}
\usepackage{graphicx}
\usepackage{amsmath}
\usepackage{amssymb}
\usepackage{bm}
\usepackage{paralist,array}
\usepackage{units} 
\graphicspath{{fig/}{./Plots/}} 
\usepackage{slashed}
\usepackage{todonotes}
\usepackage[parfill]{parskip}
\usepackage{longtable}
\usepackage{subfig}

\graphicspath{{plots/}}

\newcommand{\GeV}{\:\unit{GeV}}

\newcommand{\eV}{\:\unit{eV}}

\DeclareMathOperator{\real}{Re}
\DeclareMathOperator{\imag}{Im}

\title{Effective photon mass and (dark) photon conversion in the inhomogeneous Universe}

\author[1]{Andrés Arámburo García,}
\author[2,3]{Kyrylo Bondarenko,}
\author[4]{Sylvia Ploeckinger,}
\author[5]{Josef Pradler,}
\author[5]{and Anastasia Sokolenko}

\affiliation[1]{Leiden Observatory, Leiden University, PO Box 9513, NL-2300 RA Leiden, the Netherlands}
\affiliation[2]{Theoretical Physics Department, CERN, 1 Esplanade des Particules, Geneva 23, CH-1211, Switzerland}
\affiliation[3]{L'Ecole polytechnique f\'ed\'erale de Lausanne, Route Cantonale, 1015 Lausanne, Switzerland}
\affiliation[4]{Lorentz Institute for theoretical physics, Leiden University, Niels Bohrweg 2, Leiden, NL-2333 CA, the Netherlands}
\affiliation[5]{Institute of High Energy Physics, Austrian Academy of Sciences, Nikolsdorfergasse 18, 1050 Vienna, Austria}

\emailAdd{aramburo@strw.leidenuniv.nl}
\emailAdd{kyrylo.bondarenko@cern.ch}
\emailAdd{ploeckinger@lorentz.leidenuniv.nl}
\emailAdd{josef.pradler@oeaw.ac.at}
\emailAdd{anastasia.sokolenko@oeaw.ac.at}

\begin{document}

\abstract{
Photons traveling cosmological distances through the inhomogeneous Universe experience a great variation in their in-medium induced effective mass. 
Using the \textsc{eagle} suite of hydrodynamical simulations, we infer the free electron distribution and thereby the effective photon mass after reionization. We use this data to study the  inter-conversion of kinetically mixed photons and dark photons, which may occur at a great number of resonance redshifts, and obtain the ``optical depth'' against conversion along random lines-of-sight. Using COBE/FIRAS, Planck, and SPT measurements, we constrain the dark photon parameter space from the depletion of CMB photons into dark photons that causes both spectral distortions and additional anisotropies in the CMB. 
Finally, we also consider the conversion of \mbox{sub-eV}  dark radiation into ordinary photons. We make the line-of-sight distributions of both, free electrons and dark matter, publicly available.
}

\maketitle

\section{Introduction}

The dispersive properties of photons change inside media, a phenomenon generally referred to as refraction. If the medium is locally 
isotropic, the propagation of transversely polarized photons may be understood as them carrying an effective mass~\cite{BornWolf}.
The latter depends on the plasma frequency which in turn is principally determined by the density of free electrons,%
\footnote{See Appendix~\ref{sec:effective-mass} for discussion on corrections to (\ref{eq:mA}).}
\begin{equation}
     m_{A}(n_e) \approx \sqrt{\frac{4\pi \alpha n_e}{m_e}}.
     \label{eq:mA}
\end{equation}
Here $n_e$ is the free electron number density, $\alpha$ is the fine-structure constant, and $m_e$ is the electron mass.

The Universe is filled with free electrons, and for the better part of its history it resides in a highly ionized state. Hence, whenever the photon frequency $\omega $ approaches  (\ref{eq:mA}) or falls below, medium effects in the propagation of photons cannot be neglected. In the early Universe (for redshift $z\gtrsim 100$) the electron number density is largely homogeneous and is well-described by its spatial average value $\langle n_e\rangle$. However, at lower redshifts, inhomogeneities become large and structure formation enters the non-linear phase.
In the past two decades, numerical simulations have allowed to build the bridge between linear and non-linear regimes of gravitational collapse and to gain a detailed understanding of the dynamics of structure formation, from cosmological scales down to the scales of galactic astrophysics.
Properties of the structures in the Universe at largest scales are captured well by dark matter (DM)-only simulations~\cite{2010MNRAS.408.2163A} as it is the dominant component of matter in the Universe. However, at smaller scales, the feedback from baryonic matter is important (see e.g.~\cite{Debackere2020}). Both components can influence each other and must therefore be evolved self-consistently.

Hydrodynamical simulations such as the Evolution and Assembly of GaLaxies and their Environments (\textsc{eagle}) suite have been devised to capture such baryonic feedback processes using sub-grid physics inputs for radiative cooling, star formation, stellar mass loss, energy feedback from star formation, AGN feedback and so forth~\cite{Schaye2015, Crain2015}. Extracted quantities such as galaxy stellar mass function, star formation rates, Tully-Fisher relation, total stellar luminosities of galaxy clusters can then be compared to observations.  In this paper we use a cosmological representative volume $(100\text{ cMpc})^3$ from the \textsc{eagle} simulation to extract the number density distribution of free electrons. This simulation self-consistently evolves the baryonic profiles, and hence contains the information of $n_e$ because of overall charge neutrality.%

\begin{figure}
    \centering
    \includegraphics[width=0.7\textwidth]{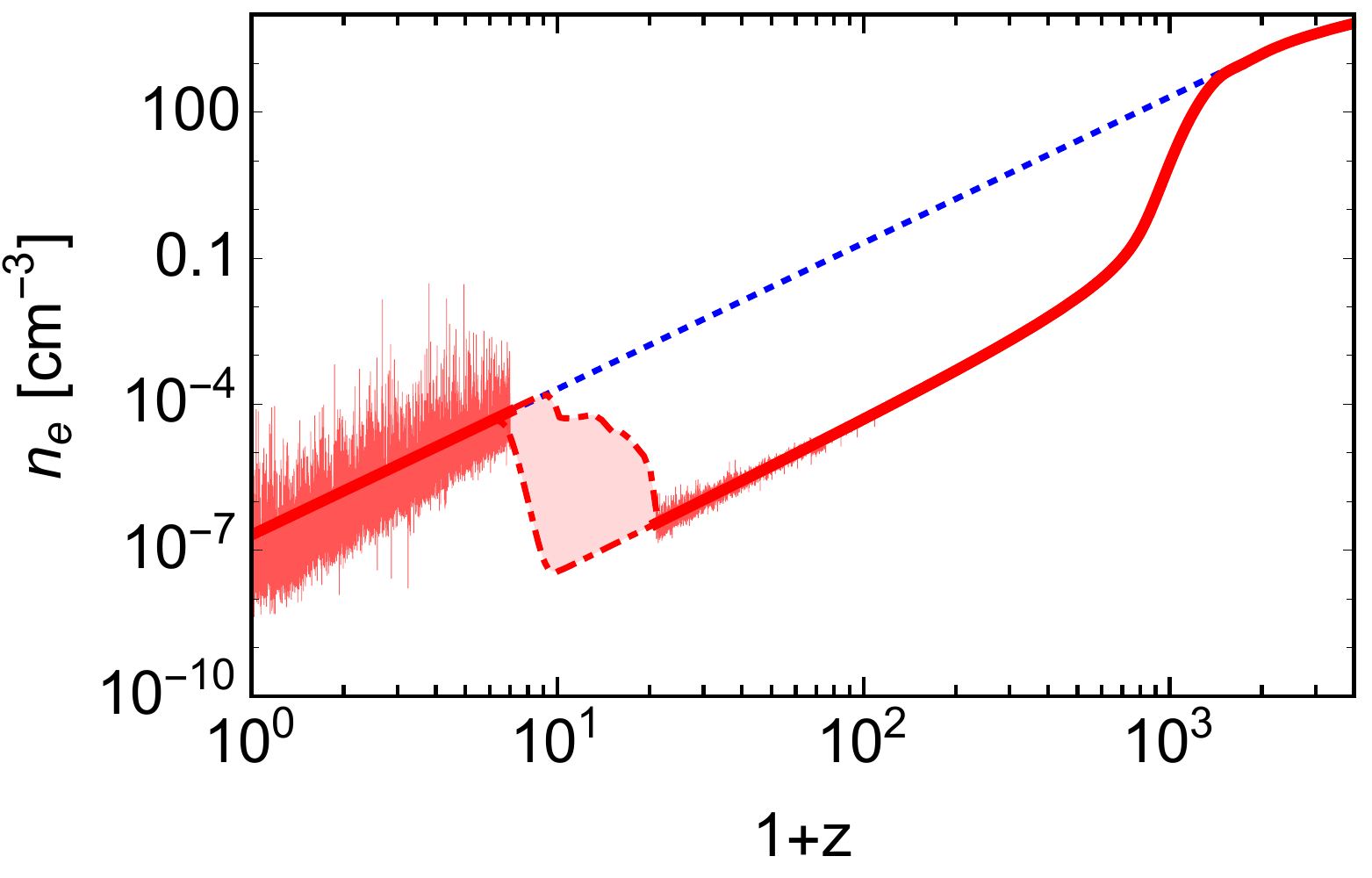}
    \caption{Free electron number density as a function of redshift (red line) and  average baryon number density (blue dotted line). The ratio between the two thick lines describes the average degree of ionisation. For $z<6$ the fluctuating electron number density along an exemplary random continuous LOS through the \textsc{eagle} simulation volume is shown. The red dot-dashed lines schematically indicate the era of reionization~\cite{Heinrich:2016ojb,Aghanim:2018eyx}, unknown in detail and not further considered in this work. Above $z>20$ we tie the fluctuations of electron density directly to the one of DM.
    }
    \label{fig:resonanses}
\end{figure}

An example for the electron density along a random continuous line-of-sight (LOS) is shown in Fig.~\ref{fig:resonanses}. The thick red (blue dotted) line shows the average density of free electrons (baryons) as a function of redshift. After recombination, at $z\sim 1100$, the average free electron fraction is $\sim 10^{-3}$ and both lines depart from each other until reionization, which concluded by $z=6$ and may have started as early as $z\lesssim 20$~\cite{Heinrich:2016ojb,Aghanim:2018eyx}. In the latter period, the average density of free electrons grows (region between dashed red lines) and the Universe transitions from mostly neutral to predominantly ionized again. This process leads to highly inhomogeneous electron density distributions, as reionization happens in patches (e.g. in the form of bubbles of ionized gas around the sources of ionizing radiation).
Post reionization, as can be seen by the spikes in Fig.~\ref{fig:resonanses}, the electron density fluctuates significantly  while following the cosmological $(1+z)^3$ trend of the mean density of the Universe. Structure formation leads to a large density contrast of the plasma in compact galaxy clusters, cosmic filaments, and volume filling voids, resulting in the highly inhomogeneous electron density distribution. 

One of the main goals of this paper is to describe the properties of the effective photon mass in the low-redshift Universe $z<6$ in a form convenient for applications.
We make the density distribution of free electrons among random LOS on the sky as well as DM density publicly available at~\cite{Zenodo}.
Also, as the effective photon mass depends on the chosen LOS we calculate the fluctuations associated with different directions on the sky and extract observables from that.

As a natural application of our results we consider models of new physics containing  dark photons that are kinetically mixed with the Standard Model photon~\cite{Okun:1982xi,Galison:1983pa,Holdom:1985ag}. In this setup both states can undergo vacuum oscillations into each other---a feature that can be constrained by the measurements of the cosmic microwave background (CMB) radiation when the dark photon mass is below the eV-scale~\cite{Georgi:1983sy,Jaeckel:2008fi}. The medium effect on photons, however, allow for a large enhancement of this conversion probability, namely, once $m_A(n_e)$ equals the dark photon mass. The mixing between both states becomes maximal, and resonant flavor conversion between the states becomes possible. Applied to the CMB, in~\cite{Mirizzi:2009iz,Kunze:2015noa,McDermott:2019lch} this was used to constrain dark photons under the assumption of a homogeneous Universe, where the electron density in (\ref{eq:mA}) was taken to be the cosmological averaged one, $\langle n_e\rangle$.%
\footnote{The cosmology and astrophysics of sub-eV dark photons has further been studied in~\cite{Nelson:2011sf,Arias:2012az,Dubovsky:2015cca,Graham:2015rva,Kovetz:2018zes,Agrawal:2018vin,Wadekar:2019xnf,AlonsoAlvarez:2019cgw}.}
The results of our simulations will allow to lift such simplifying restrictions and study dark photon-photon conversion in the cosmological context and under  realistic conditions of an inhomogeneous Universe, and extend and solidify previously derived CMB constraints.%
 \footnote{A similar approach was recently taken in~\cite{Caputo:2020bdy}. Whereas we use the direct input from simulations to extract $n_e$, \cite{Caputo:2020bdy} uses probability distribution functions (informed from numerical simulations).}
Our results can also be applied to the case in which a significant abundance of dark photons is present or is being generated as the Universe evolves, allowing for conversion from the hidden sector into the low-energy part of the CMB spectrum. A companion paper~\cite{Bondarenko:2020moh} exploits this and uses our results to  constrain the scenario put forward in~\cite{Pospelov:2018kdh} that proposes a solution of the EDGES anomaly~\cite{Bowman:2018yin} using such conversion of dark photons during the dark ages.

The paper is organized as follows:
in Section~\ref{sec:simulations} we describe the procedure of extracting of the free electron number density and DM density from simulations and of constructing a random continuous LOS thereof.
In Section~\ref{sec:model} we describe the model of a kinetically mixed dark photon and discuss properties of the resonant conversion.
In Section~\ref{sec:Conversion-CMB} we discuss the impact of conversion on the CMB spectrum and put constraints on the on the mixing between dark  and ordinary photons. In Section~\ref{sec:inverse_conversion} we consider a model where dark photons are created from DM decays and  describe the properties of the signal that is expected in this case.

\section{Description of simulations}
\label{sec:simulations}

The \textsc{eagle} simulation~\cite{Schaye2015,Crain2015} is a suite of smoothed particle hydrodynamic (SPH) simulations that follow the cosmological structure formation from $z=127$ to $z=0$ with the Planck 2013 cosmological parameters \cite{planck2013}. For this work, we use the reference simulation L100N1504 with a box size of $L=100$~comoving Mpc (cMpc). Dark matter and baryons are both modelled with $N = 1504^3$ particles with an initial particle mass of $1.81\times10^6\,\mathrm{M}_{\odot}$ for baryons and $9.7\times10^6\,\mathrm{M}_{\odot}$ for dark matter. \textsc{eagle} therefore resolves baryonic (DM) structures down to masses of $\approx 10^8\,\mathrm{M}_{\odot}$ ($\approx 10^9\,\mathrm{M}_{\odot}$).

The properties of all 6.8 billion particles in the simulation are stored at 29 discrete points in time (``snapshots''), unequally spaced in redshifts between $z=20$ and $z=0$ and publicly released~\cite{EagleDR}. For additional information on the dark matter density at $z>20$ the simulation was re-run with identical initial conditions but without hydro-dynamics until $z=20$. A full list of the snapshots used in this work can be found in Table~\ref{tab:long}.

\begin{center}
\begin{small} 
\begin{longtable}[t]{c|c|c}
\caption{Redshifts (column 1), their corresponding lookback times in Gyr (column 2) and expansion factors (column 3) of simulation outputs (``snapshots") used in this work. We do not use the \textsc{eagle} snapshots during or before reionization ($z \ge 6$) (marked with an asterisk) for the electron density, but still include them for the DM distribution. The \textsc{eagle} public data release includes snapshots for $z \le 20$ and we supplement this data with snapshots of a dark matter only simulation with the same initial conditions as in \textsc{eagle} (marked with two asterisks).}
 \label{tab:long} \\

 \multicolumn{1}{c|}{\textbf{Redshift}} & \multicolumn{1}{c|}{\textbf{Lookback time}}& \multicolumn{1}{c}{\textbf{Expansion factor}} \\ \hline 
\endfirsthead

\multicolumn{3}{c}%
{{\bfseries \tablename\ \thetable{} -- continued from previous page}} \\
\hline \multicolumn{1}{c|}{\textbf{Redshift}} & \multicolumn{1}{c|}{\textbf{Lookback time}} & \multicolumn{1}{c}{\textbf{Expansion factor}} \\ \hline 
\endhead

\hline 

\endfoot

\hline \hline

\endlastfoot

\
 0.00 & 0.00 & 1.000\\
 0.10  & 1.34 & 0.909\\
 0.18  & 2.29 & 0.846\\
 0.27  & 3.23 & 0.787\\
 0.37  & 4.16 & 0.732\\
 0.50  & 5.19 & 0.665\\
 0.62  & 6.01 & 0.619\\
 0.74  & 6.71 & 0.576\\
 0.87 & 7.37 & 0.536\\
 1.00  & 7.93 & 0.499\\
 1.26  & 8.86 & 0.443\\
 1.49 & 9.49 & 0.402\\
 1.74 & 10.05 & 0.365\\
 2.01 & 10.53 & 0.332\\
 2.24 & 10.86 & 0.309\\
 2.48 & 11.16 & 0.207\\
 3.02 & 11.66 & 0.249\\
 3.53 & 12.01 & 0.221\\
 3.98 & 12.25 & 0.201\\
 4.49 & 12.46 & 0.182 \\
 5.04 & 12.63 & 0.166\\
 5.49 & 12.75 & 0.154\\
 5.97 & 12.86 & 0.143\\
 7.05* & 13.04 & 0.124\\
 8.07* & 13.16 & 0.110\\
 8.99* & 13.25 & 0.100\\
 9.99* & 13.32 & 0.091\\
 15.13* & 13.53 & 0.062\\
 20.00* & 13.59 & 0.047\\
 22.50** & 13.61  & 0.042\\
 25.00** & 13.63 & 0.038\\
 30.00** & 13.66 & 0.032\\
 40.00** & 13.70 & 0.024\\
 50.00** & 13.72 & 0.019\\
 75.00** & 13.739 & 0.013\\
 100.0** & 13.748 & 0.009\\
 125.0** & 13.752 & 0.007\\

\end{longtable}
\end{small}
\end{center}

We extract random electron number density lines of sight (LOS) through the simulation boxes to calculate the conversion of CMB photons into dark photons in Section~\ref{sec:Conversion-CMB}. As \textsc{eagle} does not include radiative transfer nor neutral gas physics, the gas properties are modelled inaccurately before and during re-ionization. We therefore limit the electron number density LOS to $z<6$, after re-ionization is completed~\cite{2015MNRAS.447..499M}. At these redshifts, neutral gas contributes only at the per cent level to the energy density of the Universe~\cite{Fukugita:2004ee}. 

In Section~\ref{sec:inverse_conversion} we explore the signal expected if dark photons are sourced through dark matter decay and later resonantly convert into photons. For this, we extract dark matter LOS for all snapshots listed in Table~\ref{tab:long}, including the high-redshift ($z>6$) snapshots. 

While we assume a homogeneous electron density between recombination ($z\approx1100$) and $z=20$ in Section~\ref{sec:Conversion-CMB}, the validity of this assumption is verified by approximating the electron density distribution in the redshift range $20<z<125$ from the dark matter distribution, normalized to the mean ionization fraction from RECFAST~\cite{1999ApJ...523L...1S,Seager:1999km}. An example for this renormalized electron density LOS is illustrated in Fig~\ref{fig:resonanses} for $20<z<125$.

\begin{figure}[t]
    \centering
    \includegraphics[width=0.48\textwidth,trim={3cm 18.6cm 1cm 4cm},clip]{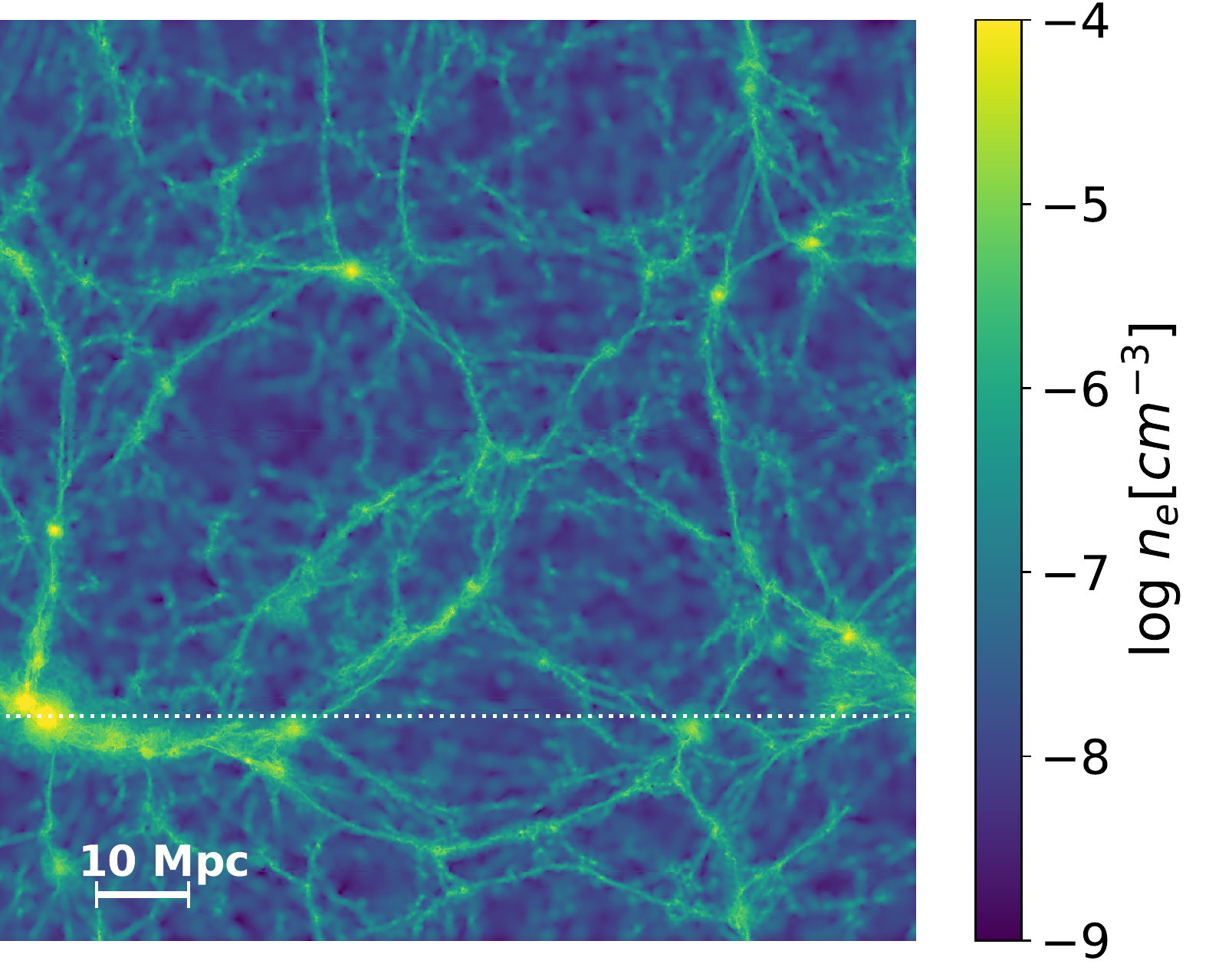}~\includegraphics[width=0.42\textwidth]{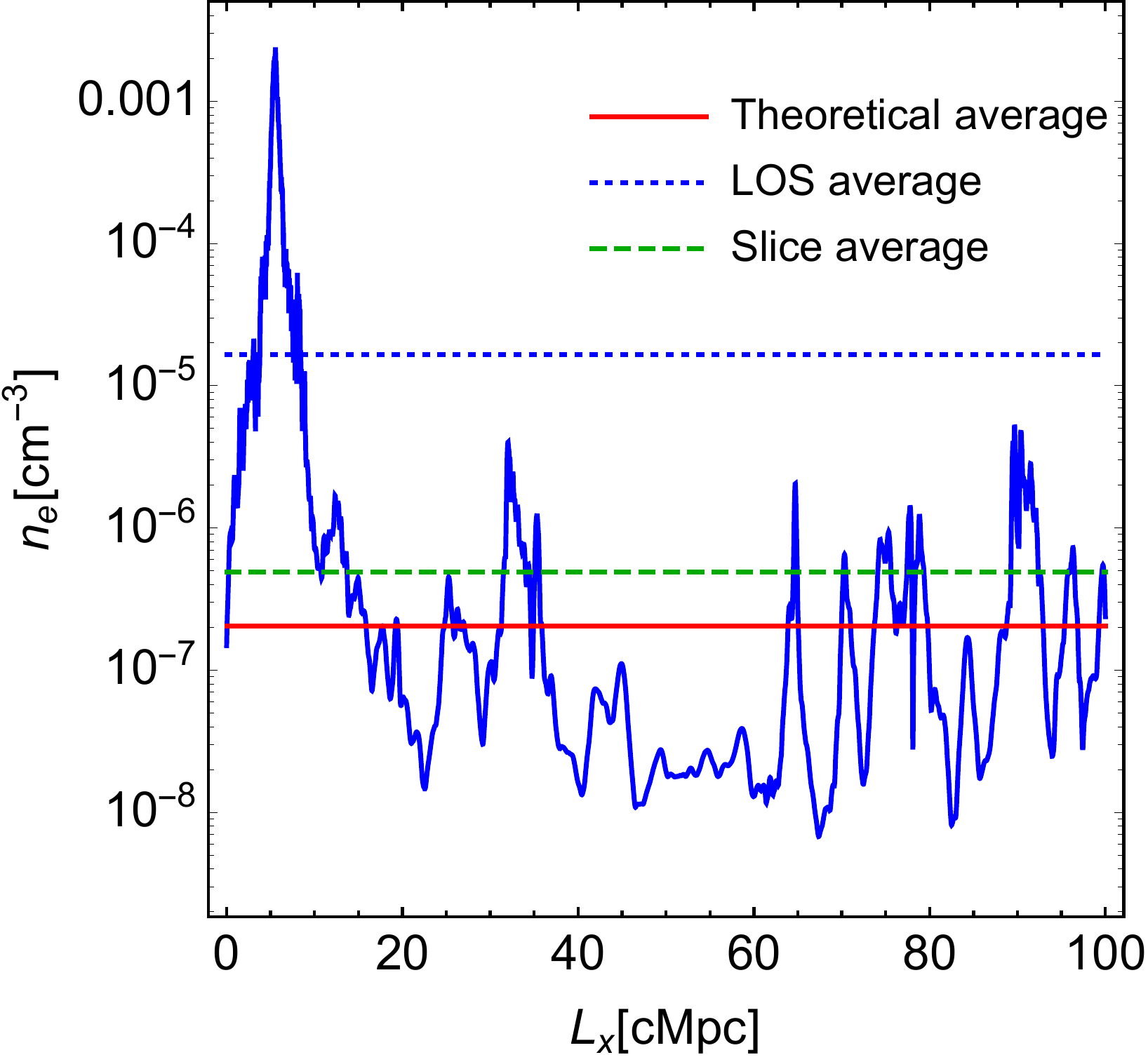}
    \caption{Electron number density image (left) at  $z=0$ with dimensions of 100 cMpc $\times$ 100 cMpc $\times$ 25 ckpc. A line of sight from the snapshot (dotted white line) is calculated in an over-dense region. The $n_e$ along this LOS is shown in the right panel, where the peaks correspond to the brighter regions in the density image, $L_x$ is a distance along the line of sight within simulation box. As this slice and LOS are selected to pass through the center of the most massive cluster of this snapshot for illustration, the average density of the LOS (``LOS average'') is higher than the average density of the full slice (``slice average''), which is still elevated compared to the mean density of the Universe (``theoretical average'').
    }
    \label{fig:ne_imgnlos}
\end{figure}

The left panel of Fig.~\ref{fig:ne_imgnlos} shows a 25~ckpc slice of the resulting electron density distribution within the simulation box at $z=0$, smoothed onto a grid with pixels of 20~ckpc$\times$20~ckpc$\times$25~ckpc\footnote{We use py-sphviewer \cite{pysph}, a public python package for fast SPH interpolation of particle properties onto a predefined grid. We also produce data for thicker LOS with pixel size 20~ckpc$\times$20~ckpc$\times$250~ckpc.
They give similar results, see Appendix~\ref{sec:thin-thick-LOS} for discussion. Later in this work we will use only thinner LOS.}. Each pixel row (or column) of the resulting electron density image corresponds to one potential line of sight. For illustration purposes the right panel of Fig.~\ref{fig:ne_imgnlos} shows the electron density along a specific LOS that passes through the center of the most massive cluster at this redshift (indicated as white dotted line in the left panel). 

The electron density is not directly available from the \textsc{eagle} data. Assuming that the contribution from neutral gas is negligible for $z<6$, the free electron number density can be calculated from the hydrogen and helium number densities, $n_{\mathrm{H}}$ and $n_{\mathrm{He}}$ and their ionization states. The contributions from the various ionization stages of other elements are negligible. The ion fractions of hydrogen and helium are calculated with the spectral synthesis code \textsc{Cloudy} v17.01~\cite{cloudy17} for gas exposed to a redshift-dependent UV background~\cite{hm12}.

Note that \textsc{eagle} does not directly model neutral gas and the neutral fraction of dense gas would have to be modeled separately. This can be done for example with fitting functions from radiative transfer simulations \cite{rahmati2013}. As we do not focus here on gas inside galaxies, this correction is not necessary.%
\footnote{
 The  cosmic fraction of neutral hydrogen after reionization is at the per-mile (per-cent) level at z=0(5.9)~\cite{Fukugita:2004ee,2015MNRAS.447..499M}. Significant amounts of HI and H$_2$ are found in the ISM where their fraction is larger than 90\% for $n_H\gtrsim 0.01\,{\rm cm}^{-3}$~\cite{Wolfire:1995fe,Wolfire:2002jm}. However, galaxies are  typically irrelevant when considering random line of sight, as we do in this work, and we are not affected by the complications of galactic astrophysics.
} 

Higher electron density peaks, such as the centres of galaxy clusters are rare and cover only a small area. The probability of a random LOS to pass through a high-density peak is therefore $\ll1$ for each snapshot. In fact, at redshift $z=0$, our simulation volume contains 10 galaxy clusters, each of which has a total mass of $>10^{14}\,\mathrm{M}_{\odot}$ and the projected area of their centres (with radius $r_c\approx 500\ \mathrm{kpc}$) cover less than 0.1 percent of the cross section of the simulation box.

\begin{figure}[t]
    \centering
    \includegraphics[width=0.43\textwidth]{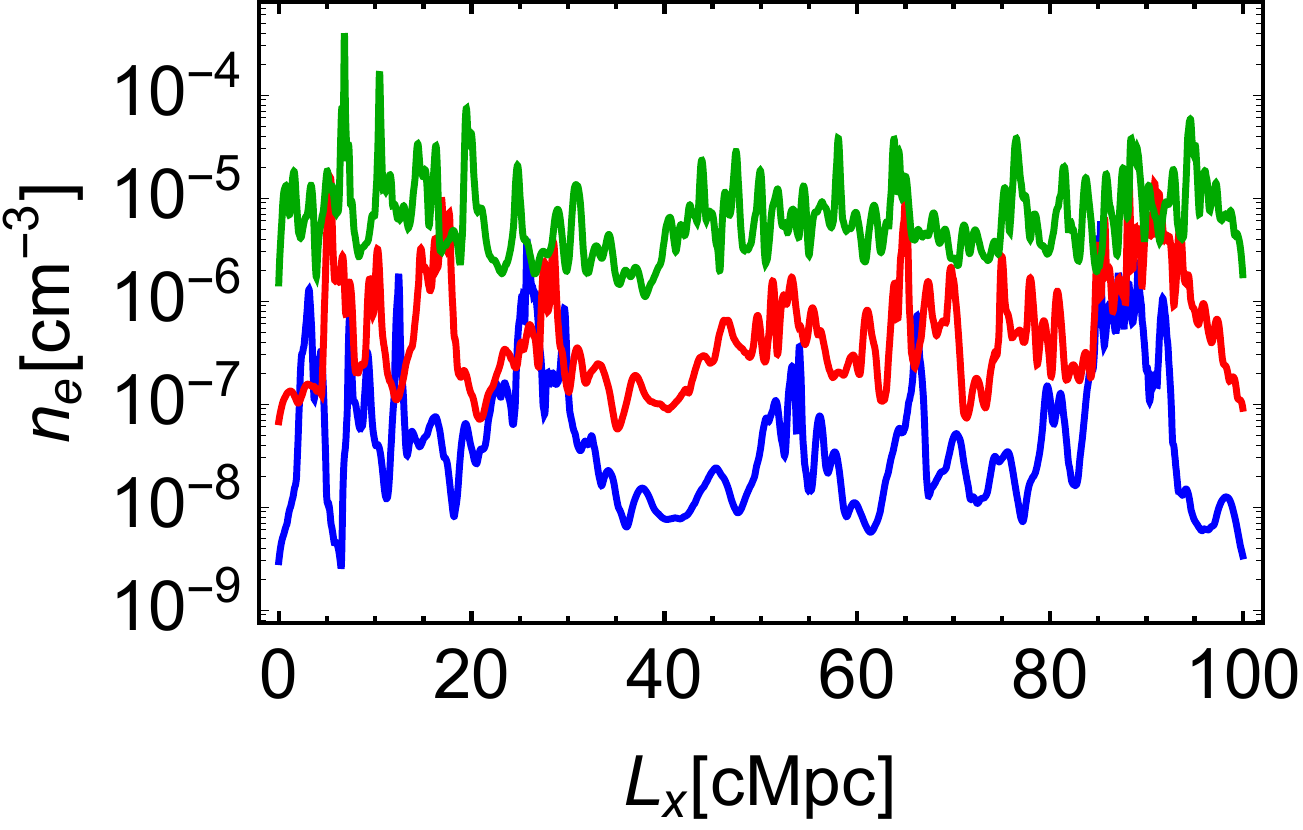}~\includegraphics[width=0.53\textwidth]{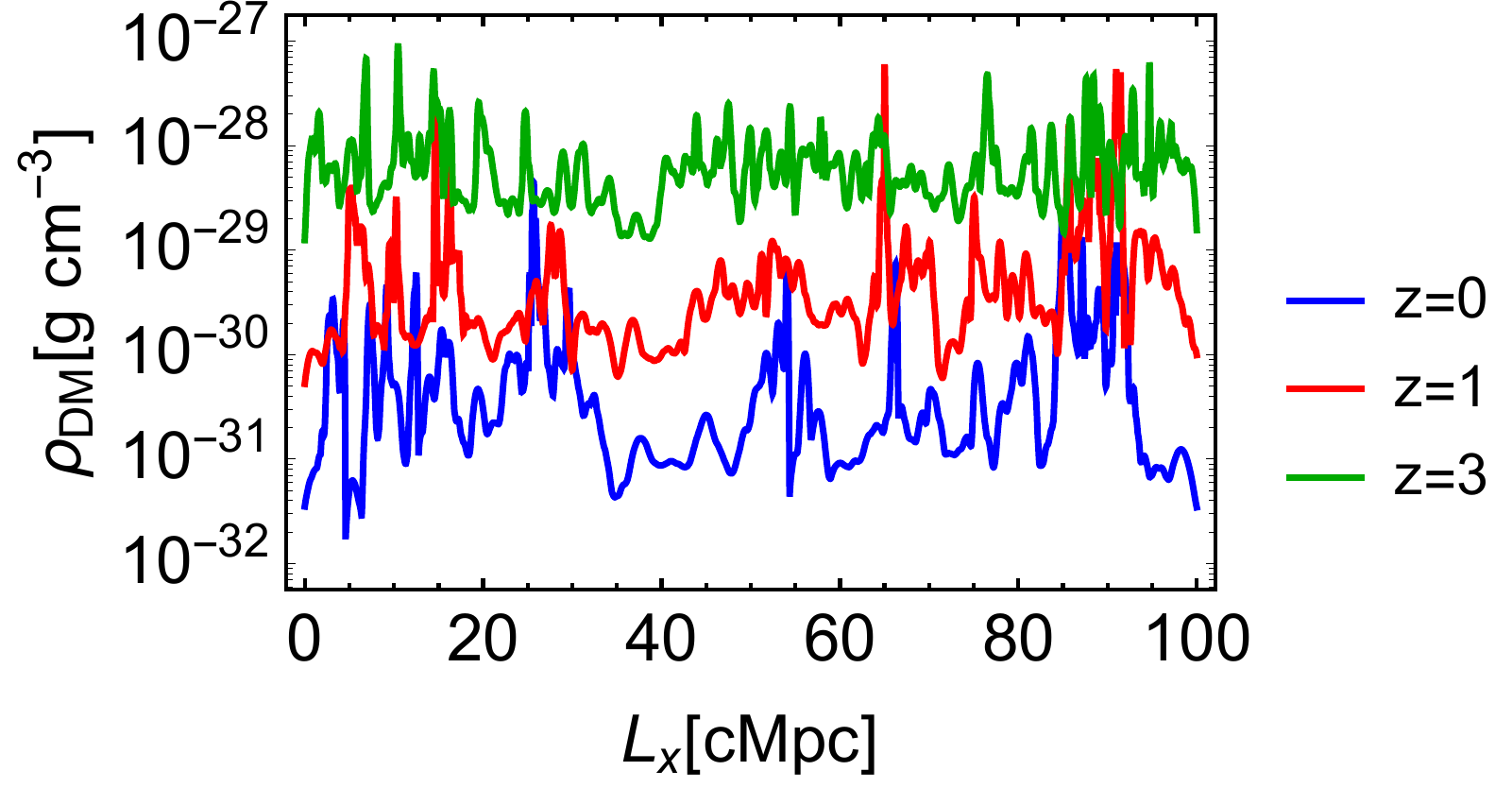}
    \caption{Electron number density (left panel) and DM density (right panel)  for the same LOS at 3 different redshifts $z=0,1,3$. $L_x$ is a distance along the line of sight within simulation box.}
    \label{fig:neEagle3LOSz}
\end{figure}

\begin{figure}[t]
    \centering
    \includegraphics[width=0.48\textwidth]{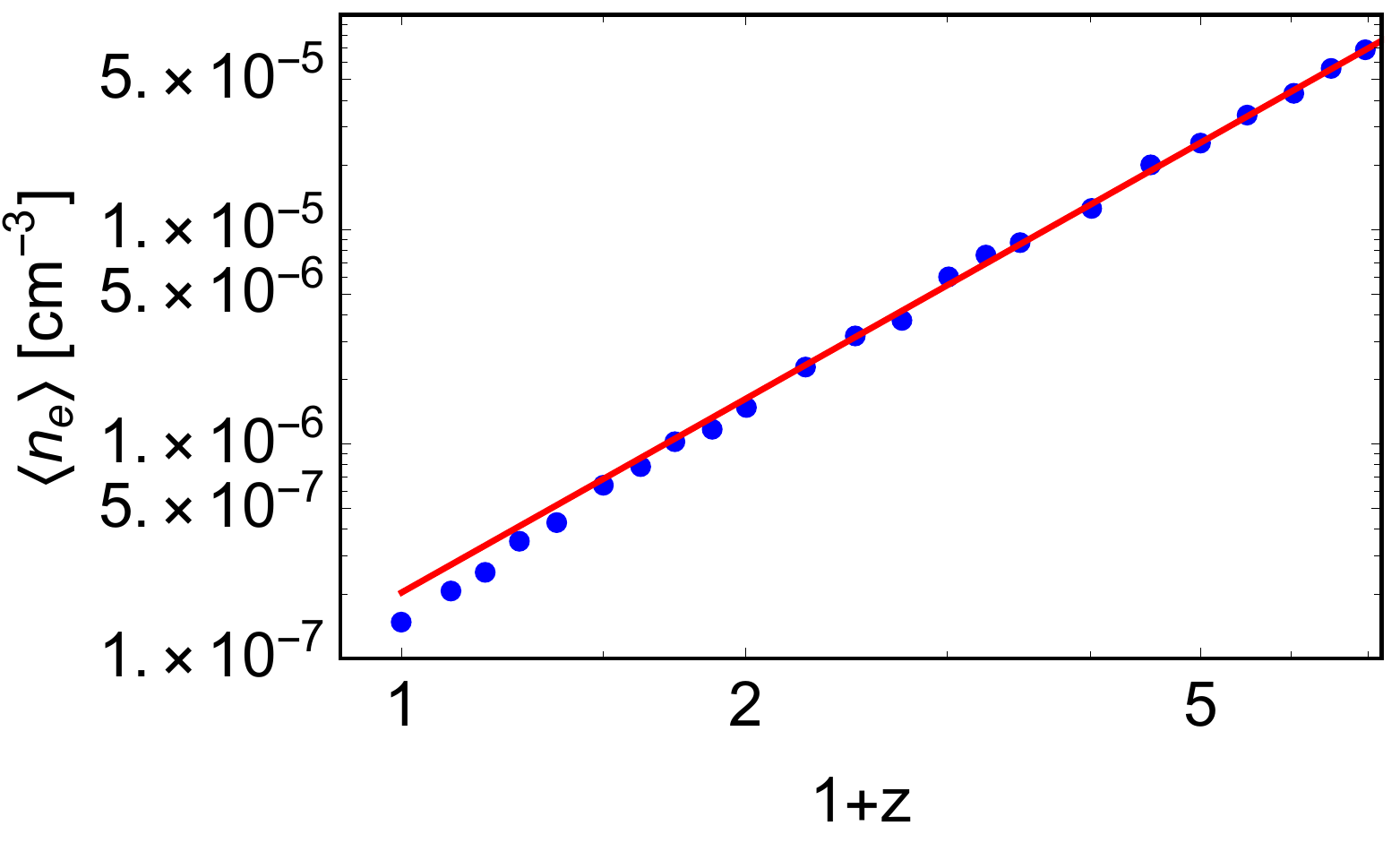}~
    \includegraphics[width=0.45\textwidth]{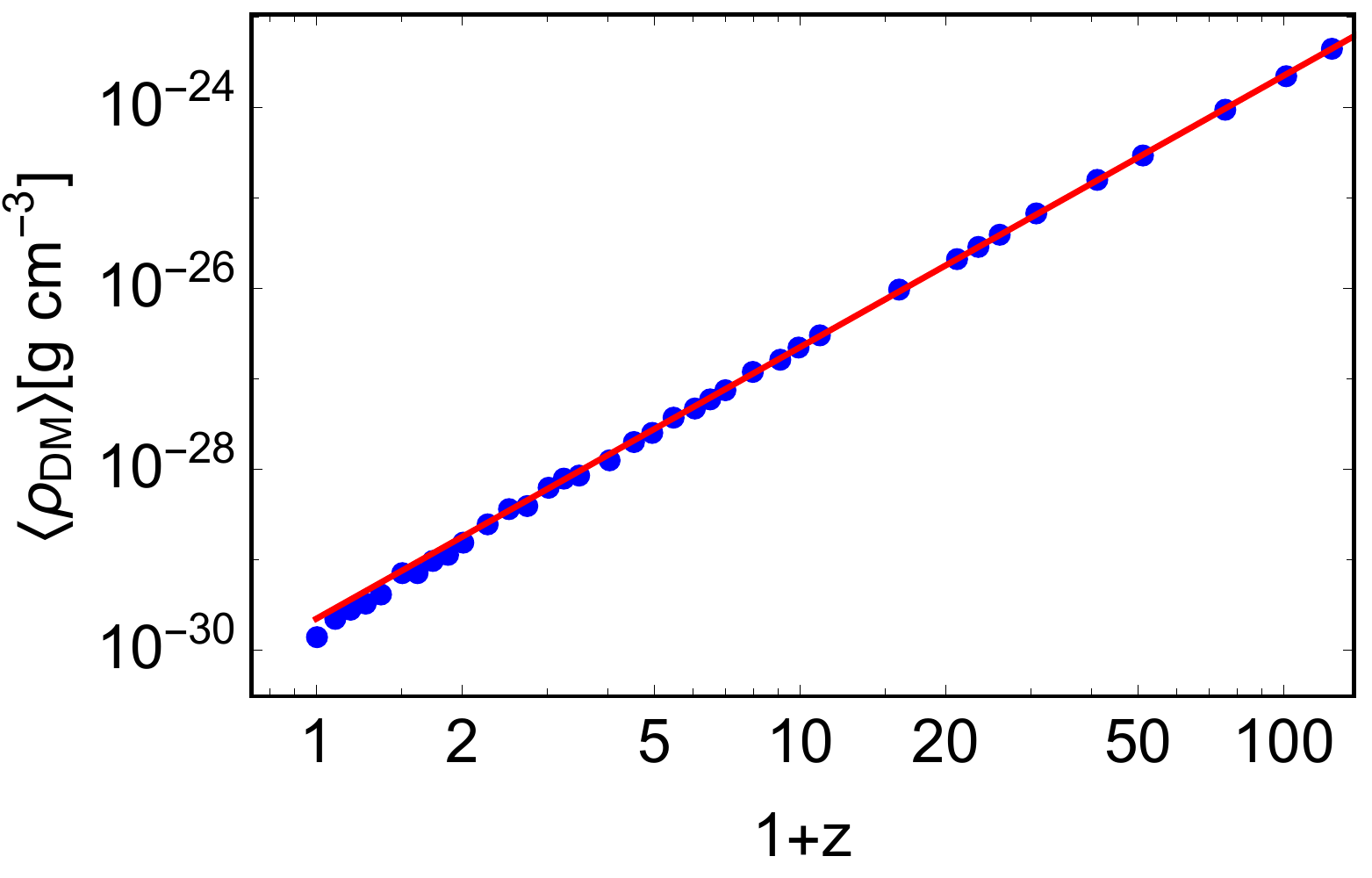}
    \caption{Mean density of electrons (left) and DM (right) averaged over 100 LOS from the snapshots listed in Table~\ref{tab:long} (blue dots);
    red lines show the cosmological mean densities.
    }
    \label{fig:neDMav}
\end{figure}

For each snapshot, 100 LOS were randomly selected, each with 5000 points along the x-axis. An example of the electron and DM number densities for the redshifts $z=0$, $z=1$ and $z=3$ is shown in Fig.~\ref{fig:neEagle3LOSz}. The average electron number density and DM density in all used snapshots ($z<6$ for $n_e$, $z\le125$ for $\rho_{\mathrm{DM}}$) is shown in Fig.~\ref{fig:neDMav}. In the left panel we see that the average values in our simulations agree well with the cosmological average in the assumption that for $z<6$ the gas is fully ionized.

\subsection{Construction of continuous lines of sight}

\begin{figure}[t]
    \centering
    \includegraphics[width=0.43\textwidth]{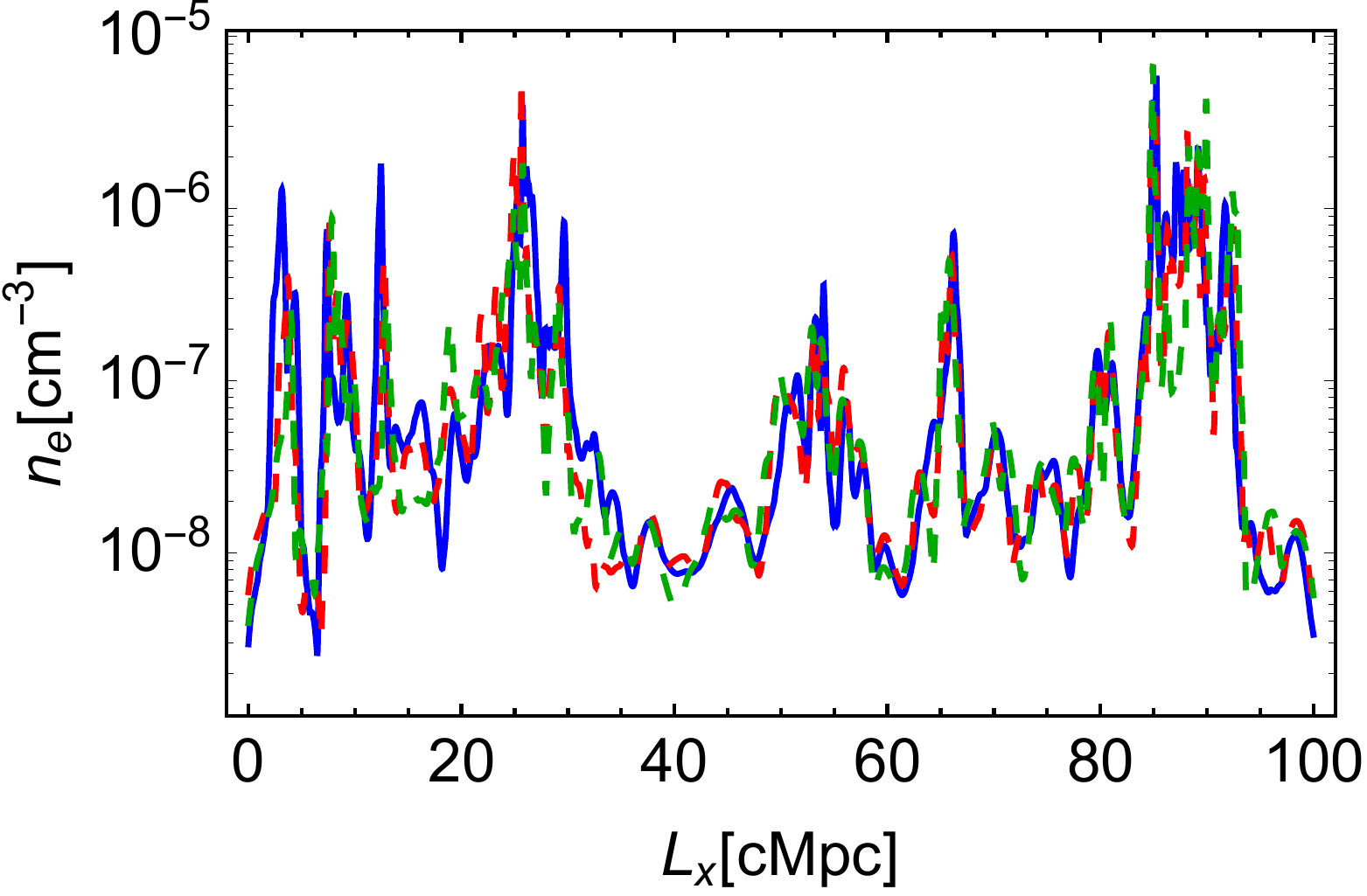}~\includegraphics[width=0.53\textwidth]{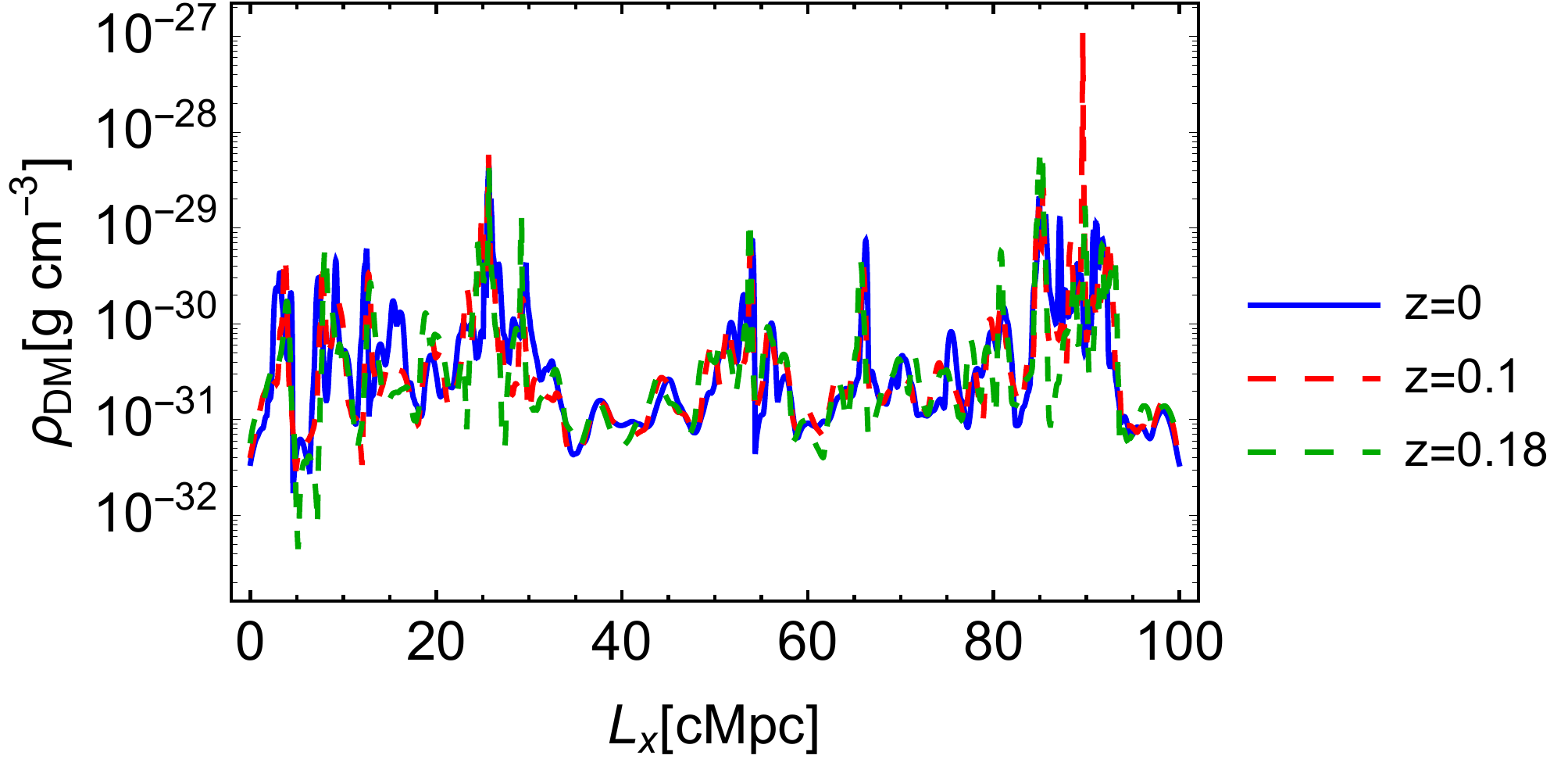}
    \caption{Electron number density (left panel) and DM density (right panel) for the same LOS from 3 neighbouring snapshots with $z=0,0.1,0.18$. The densities are scaled as $(1+z)^{-3}$, $L_x$ is a distance along the line of sight within simulation box.}
    \label{fig:check_validity}
\end{figure} 

In our simulation data, the total comoving distance of all available snapshot constitutes $\sim 30\%$ of the total comoving length of a line of sight. To fill the gaps between snapshots we take random lines of sight from the closest available snapshots and use them, properly re-scaled, to generate continuous LOS for $n_e(z)$ and $\rho_{\text{DM}}(z)$. The algorithm is the following:
\begin{enumerate}
    \item We divide the line of sight by boxes $(z_k, z_{k+1})$, each has the comoving length that same as a size of the simulation $L$.
    \item For each box, we find the nearest available simulation snapshot.
    \item In this simulation snapshot at redshift $z_{\text{sim}}$, we take a random LOS and multiply it by the factor 
    \begin{equation}
        n_e/\rho_{\text{DM}}(z_k) = n_e/\rho_{\text{DM}}(z_{\text{sim}})
        \left( \frac{1+z_{k}}{1 + z_{\text{sim}}}
        \right)^3.
    \end{equation}
    \item Finally, we take the transformed data and insert it between $(z_k, z_{k+1})$. In this way we generate continuous lines of sight.
\end{enumerate}
To check the validity of this procedure, we compare three nearest snapshots and checked that indeed, the density fluctuation in the neighboring snapshots scales as $(1+z)^{-3}$, see Fig.~\ref{fig:check_validity} where we see a good agreement between scaled LOS. This means that scaling of the nearest snapshot used in our procedure is a good approximation.

\begin{figure}[t]
    \centering
    \includegraphics[width=0.48\textwidth]{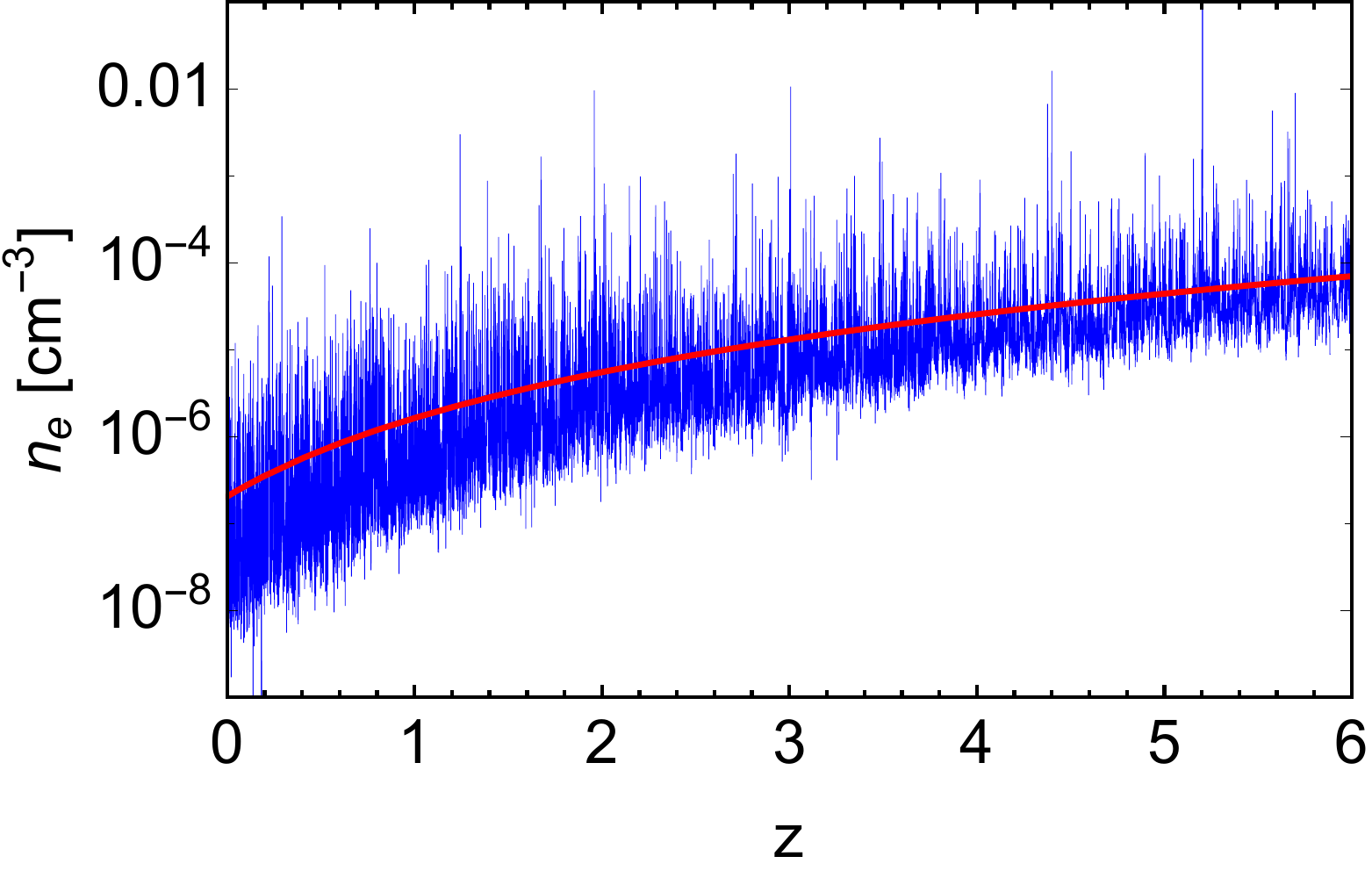}~\includegraphics[width=0.48\textwidth]{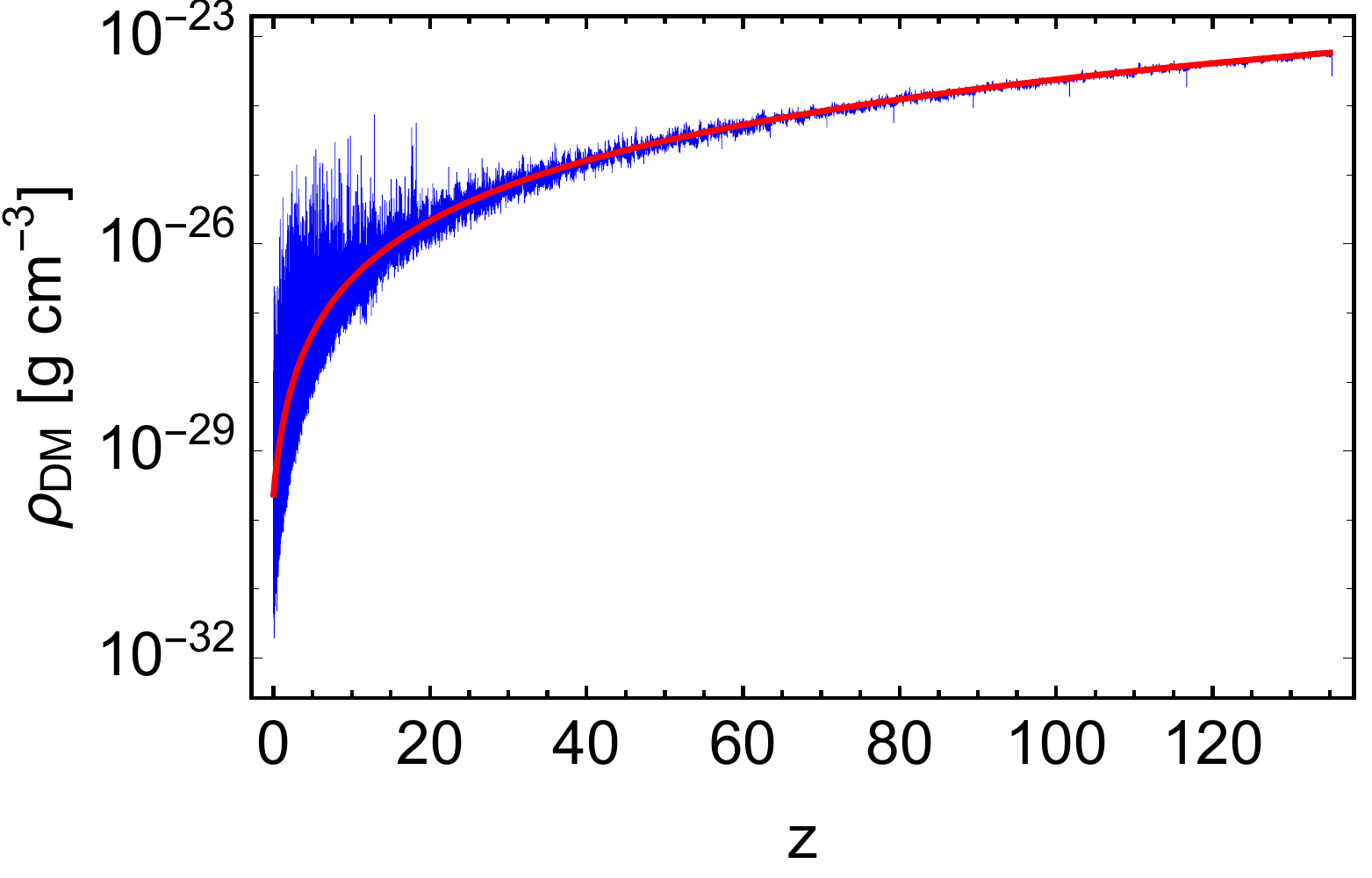}
    \caption{Electron (left) and DM (right) densities for one random continuous LOS extracted by combining all available snapshots. Red lines show the cosmological mean densities.}
    \label{fig:neDMLOS}
\end{figure}

We apply this procedure to both the electron number density $n_e$ and DM density $\rho_{\text{DM}}$. The example of the continuous LOS are shown in Fig.~\ref{fig:neDMLOS}. We make the data on the electron number density and DM density as well a code for the construction of the continuous LOS publicly available~\cite{Zenodo}.

\section{Photon - dark photon conversions}
\label{sec:model}

The prototype scenario for which the effective photon mass plays an important role is a kinetically mixed vector, or ``dark photon'' $A'$. It interacts with the photon $A$ via the kinetic mixing of their field strength tensors $F'_{\mu\nu}$ and~$F_{\mu\nu}$ with a dimensionless coupling~$\epsilon$~\cite{Okun:1982xi,Galison:1983pa,Holdom:1985ag}:
\begin{equation}
\label{L}
    \mathcal{L} =  -\frac{1}{4} F_{\mu \nu}^2 - \frac{1}{4} (F_{\mu \nu}')^2 - \frac{\epsilon}{2} F_{\mu \nu} F^{' \mu \nu} + \frac{1}{2} m_{A'}^2 (A'_{\mu})^2 + e A_\mu J_{\rm em}^\mu.
\end{equation}
Here $m_{A'}$ is the dark photon mass and the last term is the photon interaction with the electromagnetic current $J_{\rm em}^{\mu}$, which is ultimately responsible for the medium-induced mixing modification. The origin of $m_{A'}$ is phenomenologically irrelevant for the purpose of this work; it may arise from the spontaneous breaking of the $U(1)'$ or be of St\"uckelberg type.

Inside a medium, the forward scattering of photons on its constituents affects both the mixing between transversely polarized states $A$ and $A'$ as well as the photon's dispersion relation. As a result (see e.g.~\cite{An:2013yfc} for details) the mixing constant $\epsilon$ should be substituted by an effective mixing angle $\epsilon_{\rm eff}$,
\begin{align}
\epsilon_{\rm eff}^2 = \frac{\epsilon^2 m_{A'}^4}{(\real \Pi_T - m_{A'}^2)^2 + (\imag \Pi_T)^2},
\end{align}
where $\Pi_T$ is the in-medium photon self energy for transversely polarized states.
 In the case that is relevant here the dispersion relation of a photon of energy $\omega$ and momentum $\vec k$ can be written as $\omega^2 - \vec k^2 = \real \Pi_T \simeq m_{A}^2(z,n_e)$ where $m_A$ is the effective photon mass.
As can be seen, at $m_{A}^2(z,n_e) = m_{A'}^2$ the effective mixing becomes maximal and resonant conversion is possible.
The probability of $A'\to A$ conversion is then  $P_{A'\to A}(\omega) = 1 -  p$~\cite{Mirizzi:2009iz} where $p = \exp[- \pi \epsilon^2 m_{A'}^2 R/\omega]$  is the level-crossing probability of the two-state system with scale parameter $R$; $p=0$ for adiabatic transition and $p\to 1$ for non-adiabatic transitions.
In the non-adiabatic regime the conversion probability is given by
\begin{equation}
    P_{A' \to A} = P_{A \to A'} \simeq \epsilon^2 \frac{\pi m_{A'}^2}{\omega} R ,
    \label{eq:P-conv-Josef}
\end{equation}
where $\omega$ is an angular frequency of the photon at the time of conversion. Whenever $P_{A' \to A}$ reaches unity, one should use the full expression for the transition probability, however in this work we always stay in the regime $P_{A' \to A} \ll 1$.

While a photon (dark photon) propagates through the Universe, the condition of resonant conversion $m^2_A(z) = m^2_{A'}$ can be satisfied many times. 
The total probability of conversion $P_{\text{tot}}$ along some continuous LOS has contributions from a potentially large number of resonances. 
Let us denote redshifts of conversion by $z_i$ and the probability of conversion by $P_i$. 
As long as the individual probabilities $P_i\ll 1$, we can express the total conversion probability as
\begin{align}
     P_{\text{tot}}(\omega) & = \sum_i P_i - \sum_{i \neq j} P_i P_j + \dots  \simeq
    \frac{\pi \epsilon^2 m_{A'}^2}{\omega} \sum_i \frac{R_i}{(1+z_{i})} \theta(z_{\rm max} - z_i)
    \label{eq:ptot}
\end{align}
where any terms that are of higher order in $P_i$ correspond to back-and-forth conversion of quanta yielding small corrections to the overall probability which have been neglected in the last equality as they are of order $\epsilon^4$; for later convenience we have also introduced a cutoff redshift $z_{\rm max}$ that corresponds to the redshift where the initial photon (dark photon) was created. 
The scale parameter is given by~\cite{Mirizzi:2009iz}
\begin{equation}
    R_i = \left|\frac{d \ln m_A^2}{d \ell}\right|^{-1}_{\ell = \ell(z_i)} = \left|\frac{d \ln n_e}{d \ell}\right|^{-1}_{\ell = \ell(z_i)},
    \label{eq:Ri}
\end{equation}
where $\ell$ is the distance traveled by a photon (dark photon) along the line of sight and we used Eq.~\eqref{eq:mA} in the second equality.

It is clear that $P_{\text{tot}}$ depends on the distribution of matter encountered along a particular continuous LOS. Therefore, $P_{\text{tot}}$ (and any signal that is produced from conversion) will have an anisotropy that is reflective of the variance of $P_{\text{tot}}$ as one scans through (simulated) directions on the sky. This effect is important during the reionization that happens in the redshift interval $6\lesssim z \lesssim 20$~\cite{Heinrich:2016ojb,Aghanim:2018eyx}, for which the resonance 
condition can re-occur as $\langle X_e\rangle $ changes from $10^{-3}$ to practically unity. In fact, because of its patchy nature,  inhomogeneities in $n_e$ may be significant during reionization. However, our knowledge about reionization history is quite uncertain, and we exclude this region from our analysis, resulting in a conservative estimation of the overall conversion probability. Our principal investigations start with redshift $z=6$, below which the resonant density may be
achieved at a great number of redshifts $z_i$ along each continuous LOS
(see Fig.~\ref{fig:neDMLOS}) because of structure formation.  

\begin{figure}[t]
    \centering
    \includegraphics[width=0.48\textwidth]{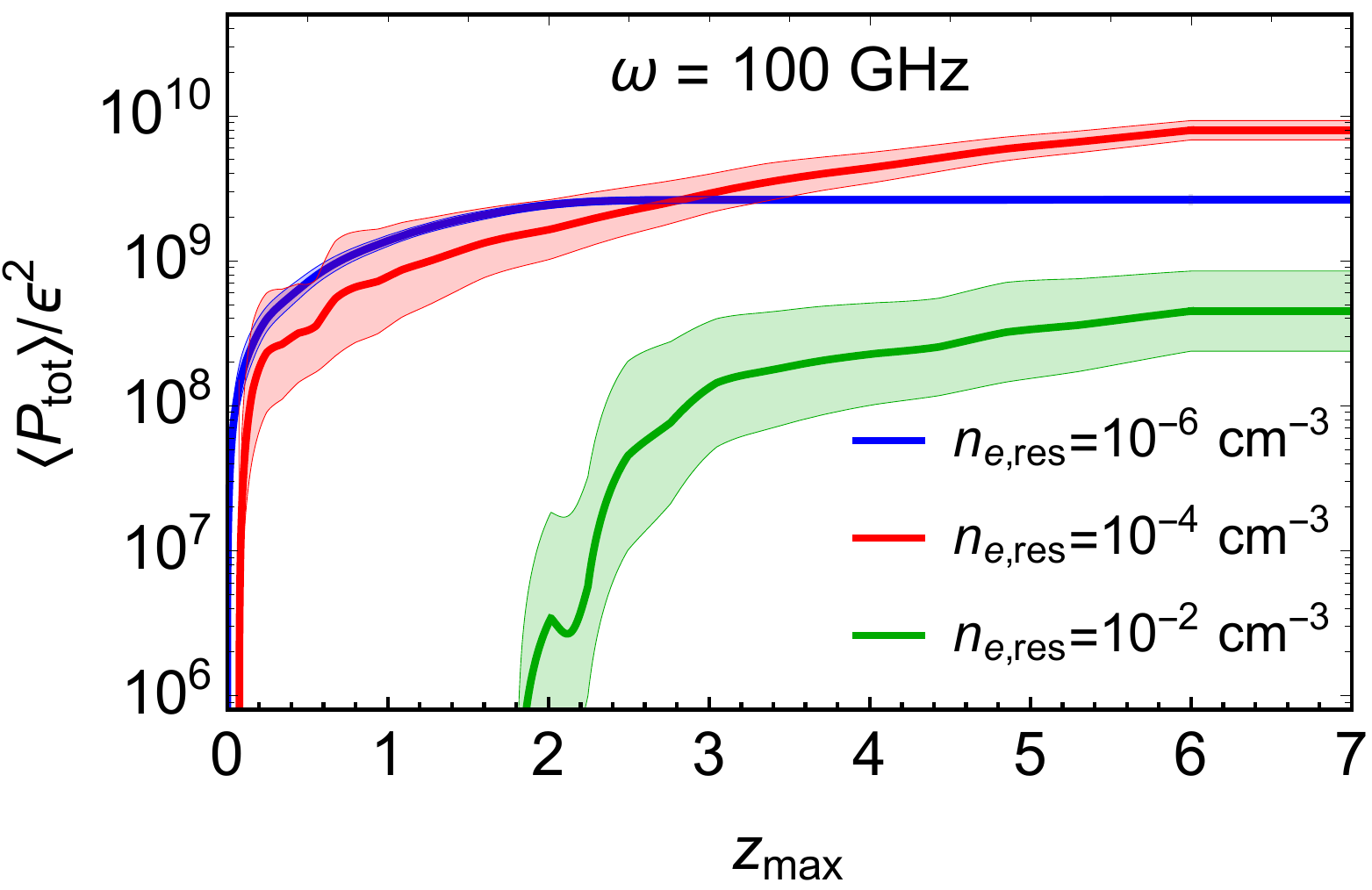}~\includegraphics[width=0.48\textwidth]{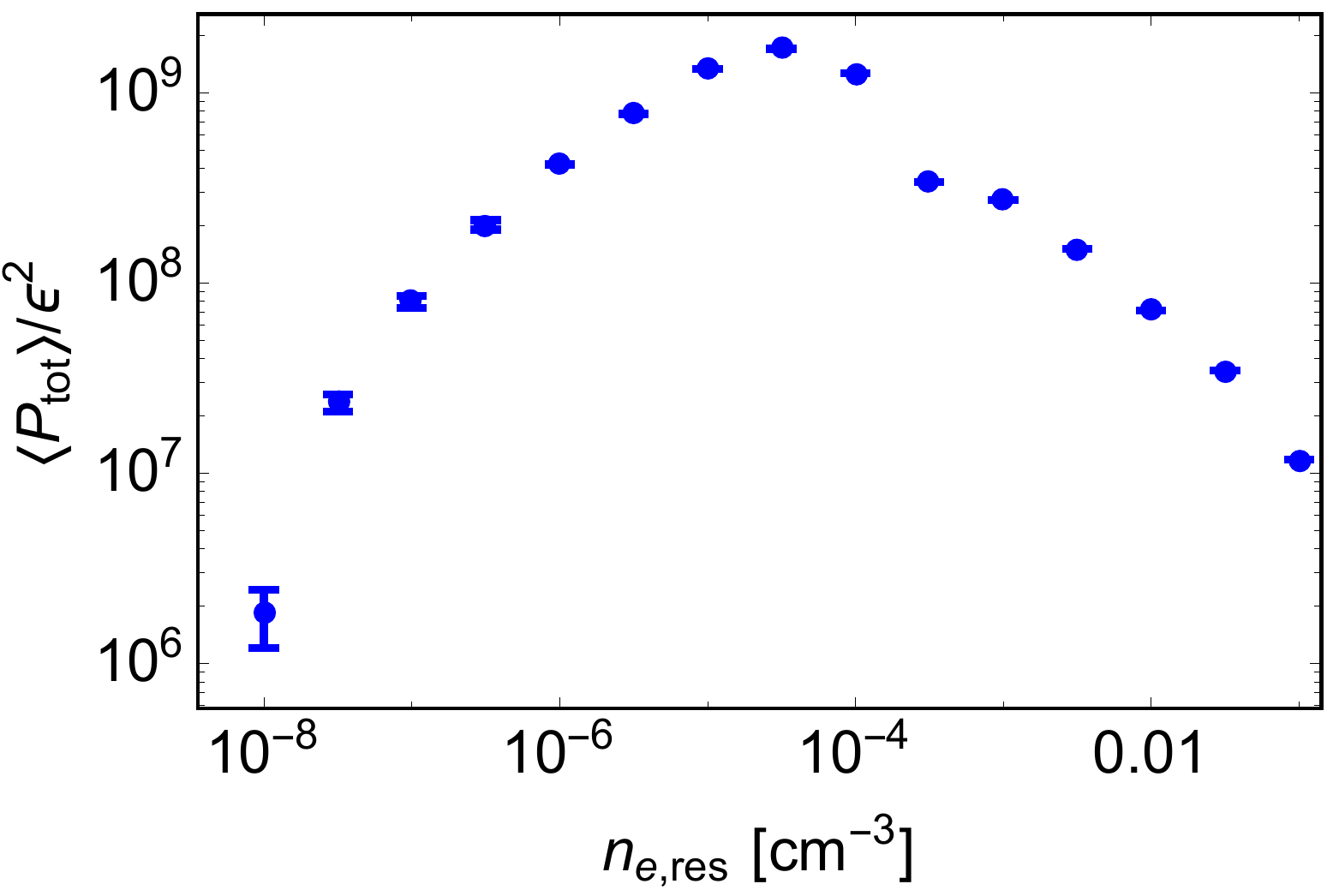}
    \caption{Average conversion probability (divided by $\epsilon^2$) as a function of $z_{\text{max}}$ (left panel) and the resonant electron number density $n_{e,\text{res}}$ (right panel). The photon frequency at Earth is fixed as $\omega = 100\text{ GHz}$, calculated using 100 random continuous LOS in simulation.
    For the right panel $z_{\text{max}} = 6$. The shaded regions on the left panel and error bars on the right panel show the scatter on this quantity.}
    \label{fig:Ptot}
\end{figure}

\begin{figure}[t]
    \centering
    \includegraphics[width=0.48\textwidth]{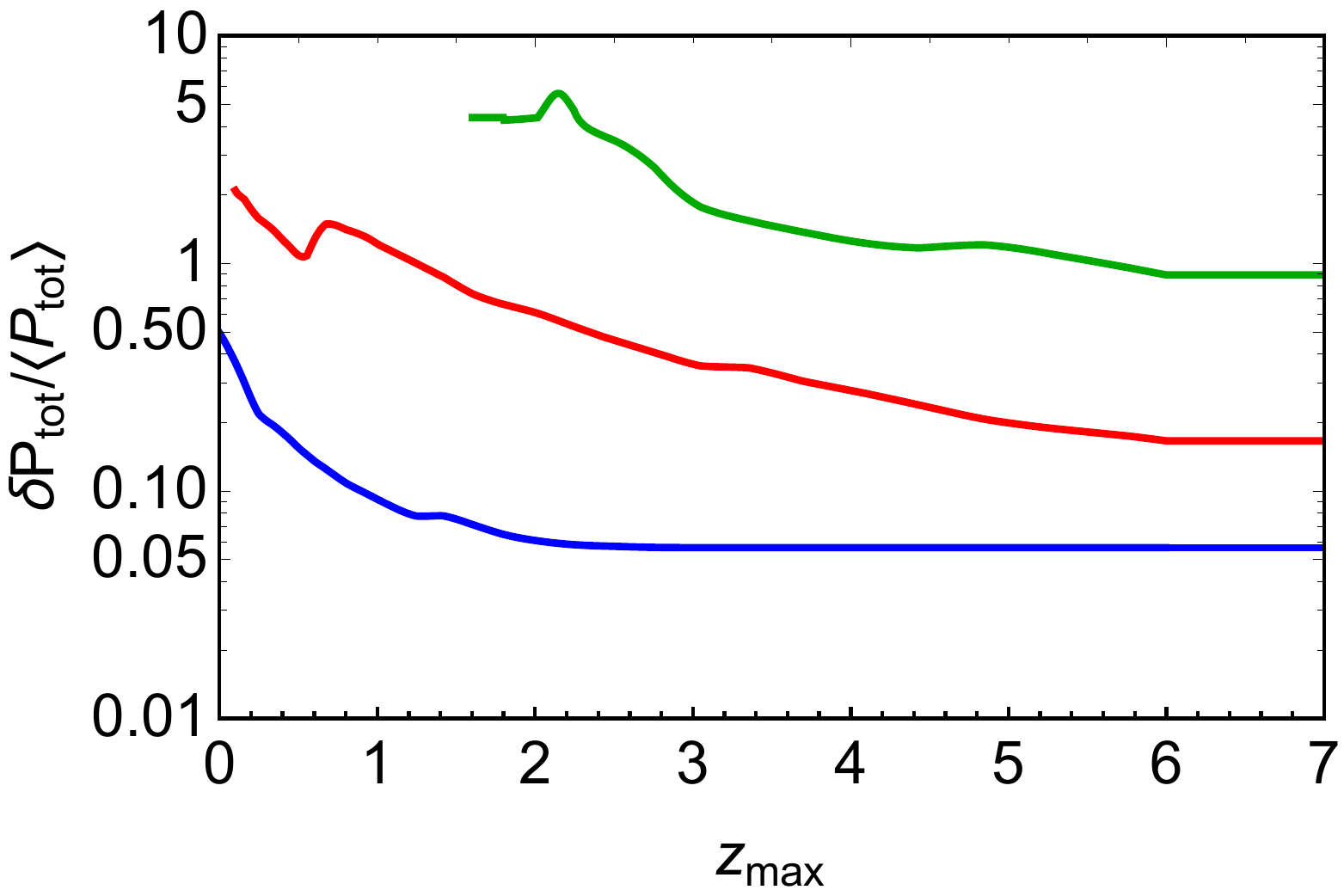}~\includegraphics[width=0.48\textwidth]{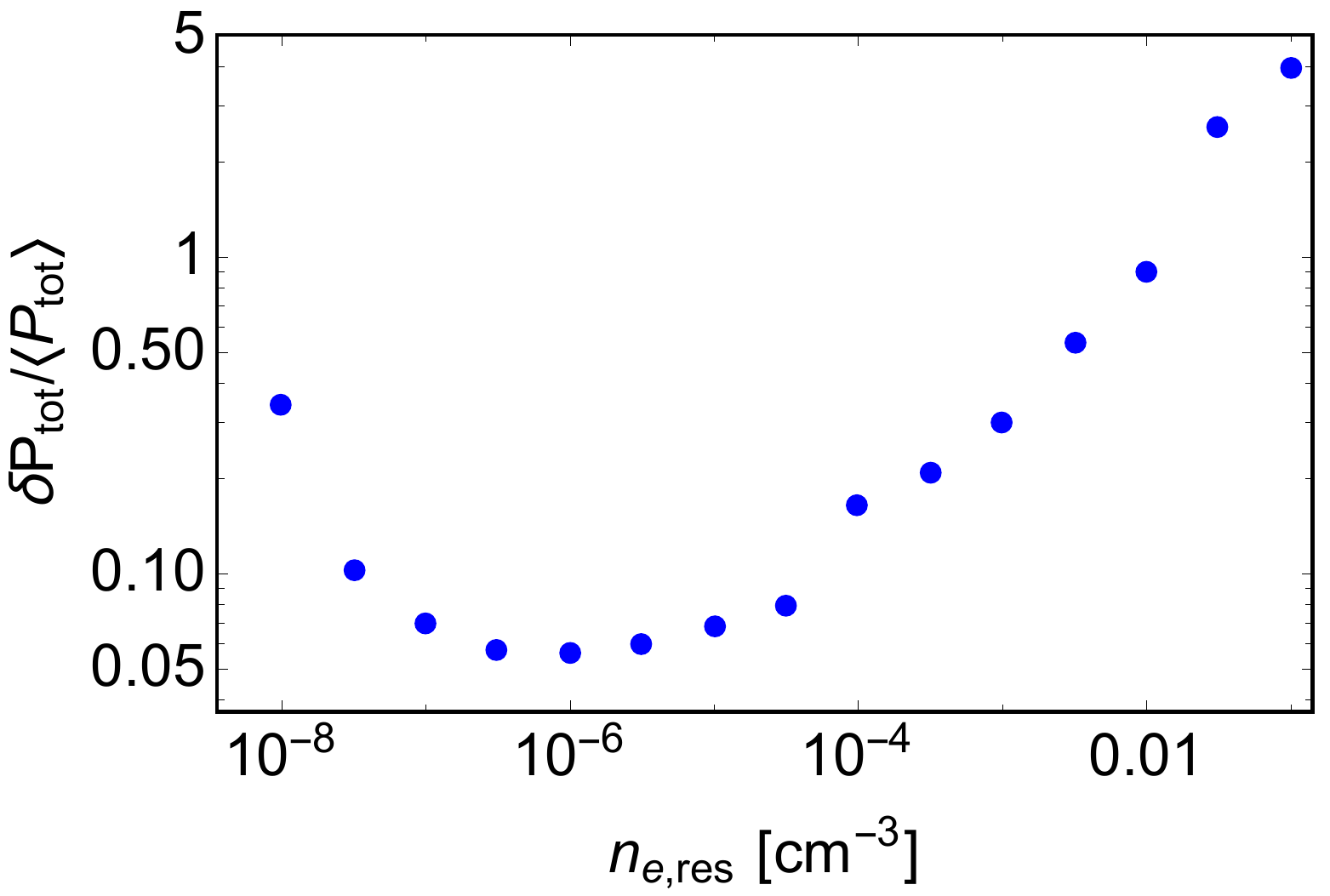}
    \caption{
    The ratio of the standard deviation and the average value of the probability of conversion, as a function of $z_{\text{max}}$ (left panel) and the resonant electron number density $n_{e,\text{res}}$ (right panel). Parameters are chosen in the same way as in Fig.~\ref{fig:Ptot}.}
    \label{fig:deltaRconv}
\end{figure}

From Eq.~\eqref{eq:ptot} follows that $P_{\text{tot}}$ is inversely proportional to the frequency $\omega$, favoring the conversion of low energy quanta,  and that it depends on the resonant concentration $n_{e,\text{res}}$ (fixed by $m_{A'}$) and redshift $z_{\text{max}}$. The average value of $P_{\text{tot}}$, obtained as the mean from 100 random continuous LOS, and divided by $\epsilon^2$ as a function of the cutoff redshift $z_{\rm max}$ and for three choices of resonant electron number density is shown in the left panel of Fig.~\ref{fig:Ptot}.\footnote{
We have checked the convergence of the averaging procedure by taking random subsets with 30, 50 and 80 continuous LOS of these 100 LOS and found that the standard deviation did not change by more than 10\%, which is sufficient for our purposes.}
As can be seen,  $\langle P_{\rm tot}\rangle$ saturates with the earliest encountered resonance either before or at redshift $z=6$ beyond which we stop the continuous LOS simulation for $n_e$. In the right panel of Fig.~\ref{fig:Ptot} we show average conversion probability as a function of the resonant electron number density for $z_{\max} = 6$. It has maximum at $n_{e,\text{res}}\sim 3\times 10^{-5}\text{ cm}^{-3}$, where we expect the largest number of resonant conversions, see Fig.~\ref{fig:neDMLOS}.

In Fig.~\ref{fig:deltaRconv} we show the standard deviation of the conversion probability, $\delta P_{\rm tot}/\langle P_{\rm tot}\rangle$. It has a minimum at the resonance number density $\sim 10^{-6}$~cm$^{-3}$, i.e.~for a  value of $m_{A'}$ where a particularly large number of resonant points along each LOS is encountered. For lower and higher resonant number density, the variance grows. The scatter in $\langle P_{\rm tot}\rangle$ will translate into anisotropies of the observed photon flux, and is always in excess of  $5\%$. This number, however, is intrinsic to the LOS size, for which we chose a cross-sectional area of $(25\times 20)\,{\rm ckpc}^2$, with a physical area that is dependent on redshift (see Section~\ref{sec:Conversion-CMB} for more detailed discussion).

\begin{figure}[t]
    \centering
    \includegraphics[width=0.6\textwidth]{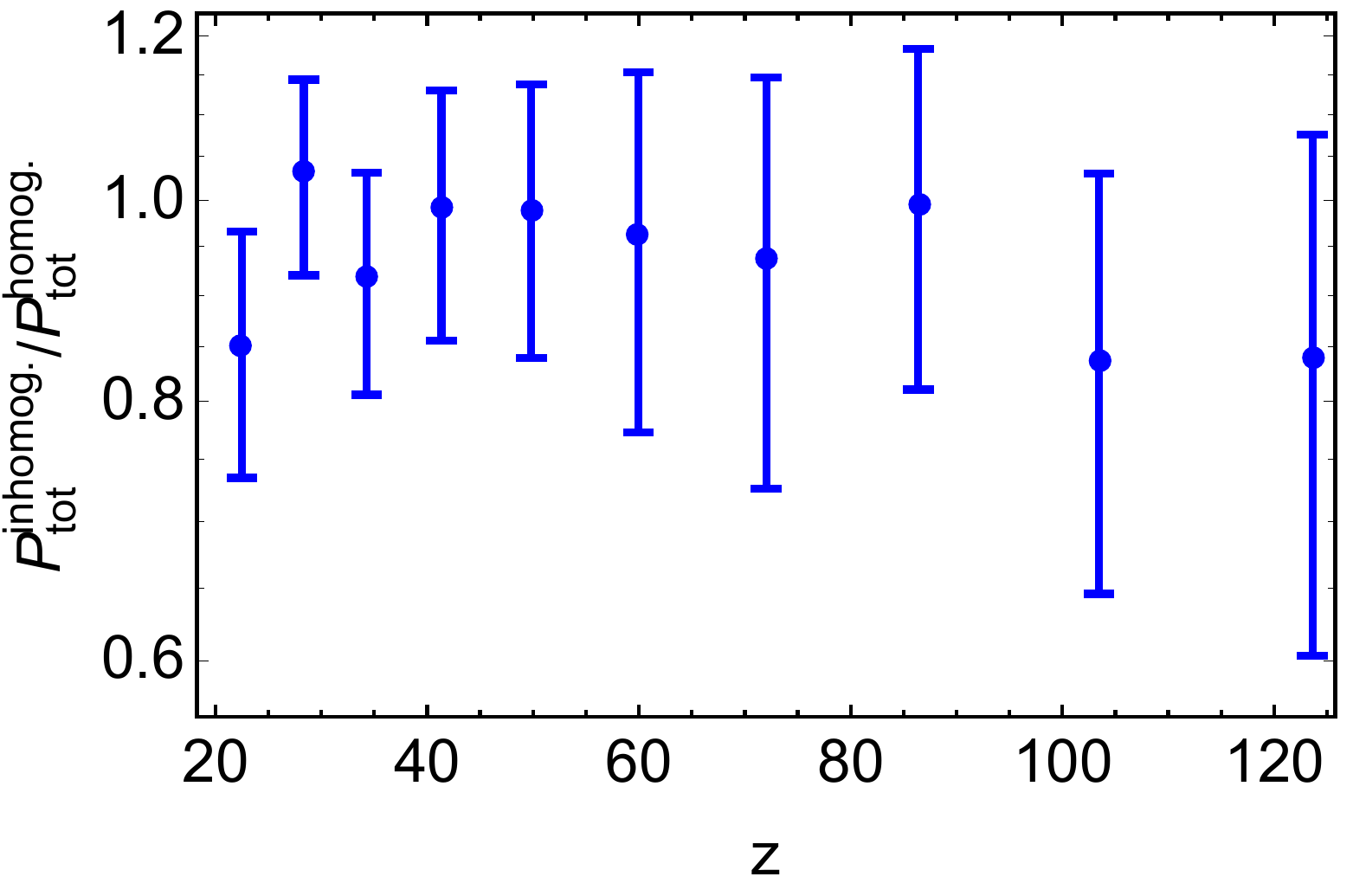}
    \caption{The ratio of conversion probabilities calculated using simulations ($P_{\rm tot}^{\rm inhomog.}$) and in the assumption of the homogeneous Universe ($P_{\rm tot}^{\rm homog.}$) for conversion in the redshift range $20<z<125$. The mean value and the variance is calculated using 100 random continuous LOS. 
    }
    \label{fig:PtotInhomoToHomo}
\end{figure}

In our simulations we also have DM data between redshifts $20$ and $125$, where the Universe was assumed homogeneous in the previous works~\cite{Mirizzi:2009iz,Kunze:2015noa}. We use scaled DM data as the electron number density (as it is shown in Fig.~\ref{fig:resonanses}) and calculate conversion conversion probability for such 100 LOS.
The result for the ratio of conversion probability including  inhomogeneities and conversion probability in the assumption of the homogeneous Universe is shown in Fig.~\ref{fig:PtotInhomoToHomo}. We see that this ratio is close to unity, so the assumption of a homogeneous conversion works well for $z>20$.

In the next sections we will use the obtained result on the conversion probability to predict spatial and spectral properties of observable signals that result from $A \leftrightarrow A'$ conversions. We consider two scenarios: i) the conversion of CMB photons into dark photons which is insensitive to a pre-existing dark photon abundance, and ii) the conversion of $A'$ fluxes into usual photons. For both cases, the total probability of conversion is given by Eq.~\eqref{eq:ptot} with an appropriate choice of $z_{\rm max}$ to be discussed below.

\section{Conversion of CMB photons into dark photons}
\label{sec:Conversion-CMB}

The resonant conversion of photons will result in distortions of the CMB spectrum, which in turn limits the photon-dark photon interaction~\cite{Mirizzi:2009iz,Kunze:2015noa}. Previous constraints were obtained using the conversion in the homogeneous limit. It was assumed that after recombination, for $z \gtrsim 20$, the density of free electrons follows the cosmological average with the resonance condition inferred from the spatially averaged version, $ m_A(z, \langle n_e(z_{\rm res}) \rangle) = m_{A'}$ at a single redshift $z_{\rm res}$.
As we discussed in the previous section, this assumption works  well in the high-redshift Universe, where it was also applied in the context of the cosmological 21~cm signal~\cite{Pospelov:2018kdh}.
Here we discuss the late-time inhomogeneous conversions that give rise not only to modifications of the CMB spectrum, but also to additional angular anisotropies in the CMB that are probed through precision cosmological observations~\cite{Bondarenko:2020moh}.

\begin{figure}[t]
    \centering
    \includegraphics[width=0.6\textwidth]{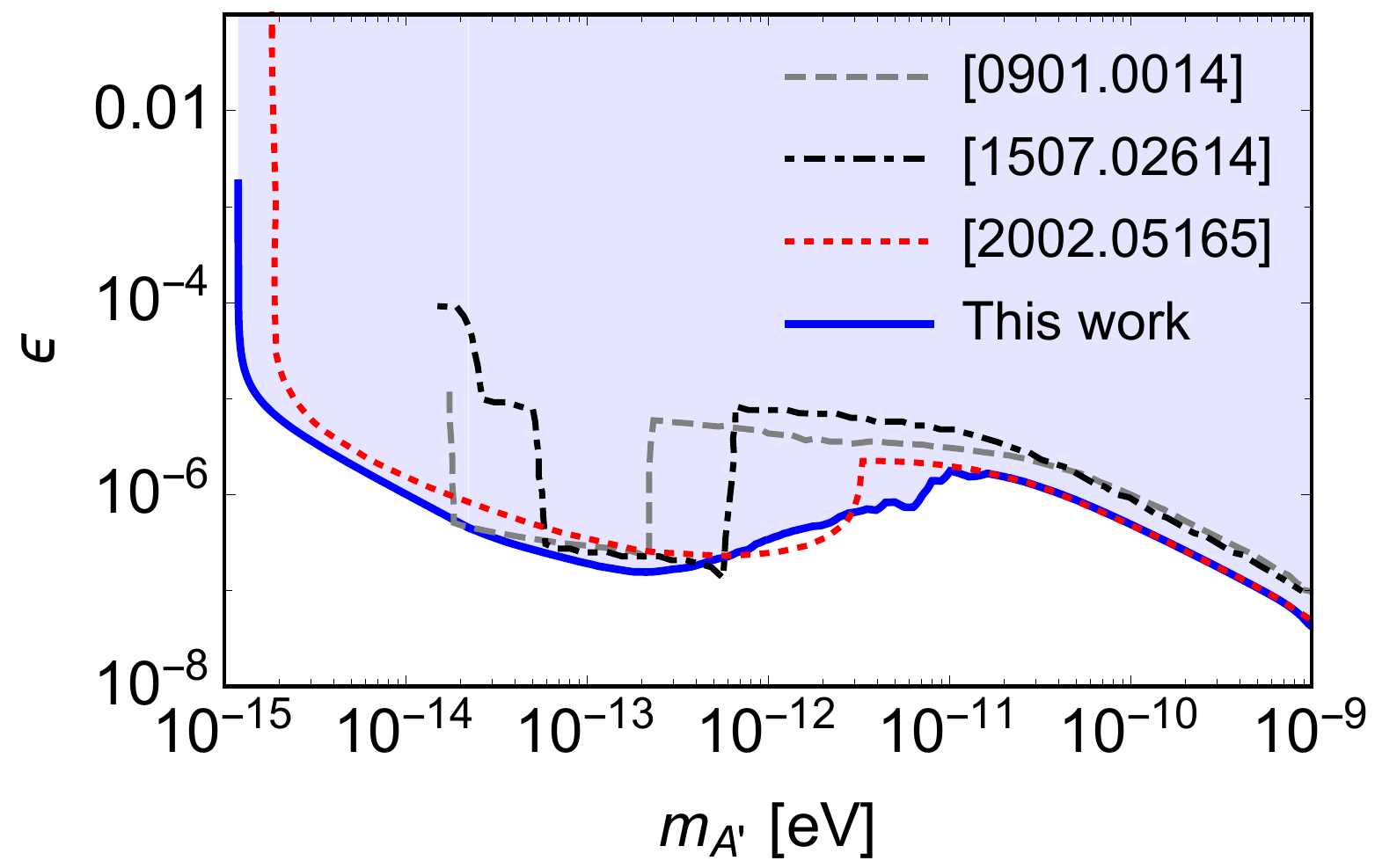}
    \caption{Constraints on the dark photon parameter space from distortions of the CMB spectrum measured by the COBE/FIRAS (blue line). The gray dashed and black dot-dashed line show the constraints obtained in~\cite{Mirizzi:2009iz,Kunze:2015noa} under the assumption of a homogeneous electron number density. The red dotted line is a result of recent~\cite{Caputo:2020bdy} that uses a semi-analytic approach to inhomgeneous conversion employing PDFs for the distribution of electron density; the results are in good agreement. The shaded region indicates the excluded parameter space. 
    }
    \label{fig:eps}
\end{figure}

The loss of CMB photons through resonant conversion induces spectral modifications and departures from the blackbody law. With its strength regulated by $\epsilon$ and its occurrence governed by $m_{A'}$ we can derive limits on the combination of those two parameters. 
Departures of the absolute flux of photons are constrained from measurements of COBE/FIRAS~\cite{Fixsen:1996nj}, which determined the CMB spectrum in the frequency range 68 to 637~GHz with a precision of $10^{-4}$.  In addition, as we have seen above, the conversion probability is anisotropic. At the end of this Section we will hence place additional constraints on excess variations of the photon flux from Planck and South Pole Telescope (SPT). 

Because of the resonant conversion some part of the CMB photons will become dark photons and will be lost, 
\begin{align}
\label{Btot}
    B_\omega(\omega) = B^{\text{CMB}}_{\omega}(\omega) [1 -  \langle P_{\text{tot}}(\omega)\rangle ],
\end{align}
where $B^{\text{CMB}}_{\omega}$ is the spectral radiance of the unmodified CMB and $P_{\text{tot}}(\omega)$ is given by~\eqref{eq:ptot} using $z_{\text{max}} = 1700$ and averaged as described above. 
We then fit Eq.~\eqref{Btot} to the COBE/FIRAS data, for which the overall temperature of the CMB is allowed to float in $B^{\text{CMB}}_{\omega}$. 

In Fig.~\ref{fig:eps} we present the excluded region of parameter space with $2\sigma$ confidence level (shaded region above blue line). For comparison, previous results obtained in~\cite{Mirizzi:2009iz,Kunze:2015noa} and~\cite{Caputo:2020bdy} are shown as labeled. The former set of papers calculated the constraint under the assumption of a homogeneous Universe, while the latter paper (red line) is the result of a recent work that included information about inhomogenities using a semi-analytic approach adopting various probability  distribution functions (PDFs) for the electron number density. Note that the difference in the gray and black lines in the region of dark photon mass from $10^{-14}$ to $6\times 10^{-13}$~eV is coming from the different assumptions about redshift of reonization considered in works~\cite{Mirizzi:2009iz,Kunze:2015noa}, see right Fig.~1 in~\cite{Mirizzi:2009iz} and Fig.~2 in~\cite{Kunze:2015noa}.
As can be seen, the main benefit of using inhomogeneous conversion is the extended reach in $m_{A'}$, corresponding to values of $n_{e,\rm res}$ that are simply not met when using cosmological average values for the electron density. For  $m_{A'} > 10^{-11}$~eV, i.e.,~$n_{e,\text{res}} \gtrsim 0.1\text{ cm}^{-3}$, there are no resonances in our simulations at low redshift, and the constraint is derived from the high-redshift conversion at $z>20$ in the homogeneous limit and with $\langle X_e \rangle$ obtained with  RECFAST~\cite{1999ApJ...523L...1S,Seager:1999km}. In the region where we set the constraint from simulations our result  agrees with~\cite{Caputo:2020bdy} in the interval $3\times 10^{-15} \leq m_{A'}/\eV\leq3 \times 10^{-12}$. In the small adjacent regions outside that interval our results differ, presumably because the method used in~\cite{Caputo:2020bdy} does not take into account the largest under- and overdensities, as they restrict their analysis to $10^{-2} < 1 + \delta < 10^2$.

\begin{figure}[t]
    \centering
    \includegraphics[width=0.6\textwidth]{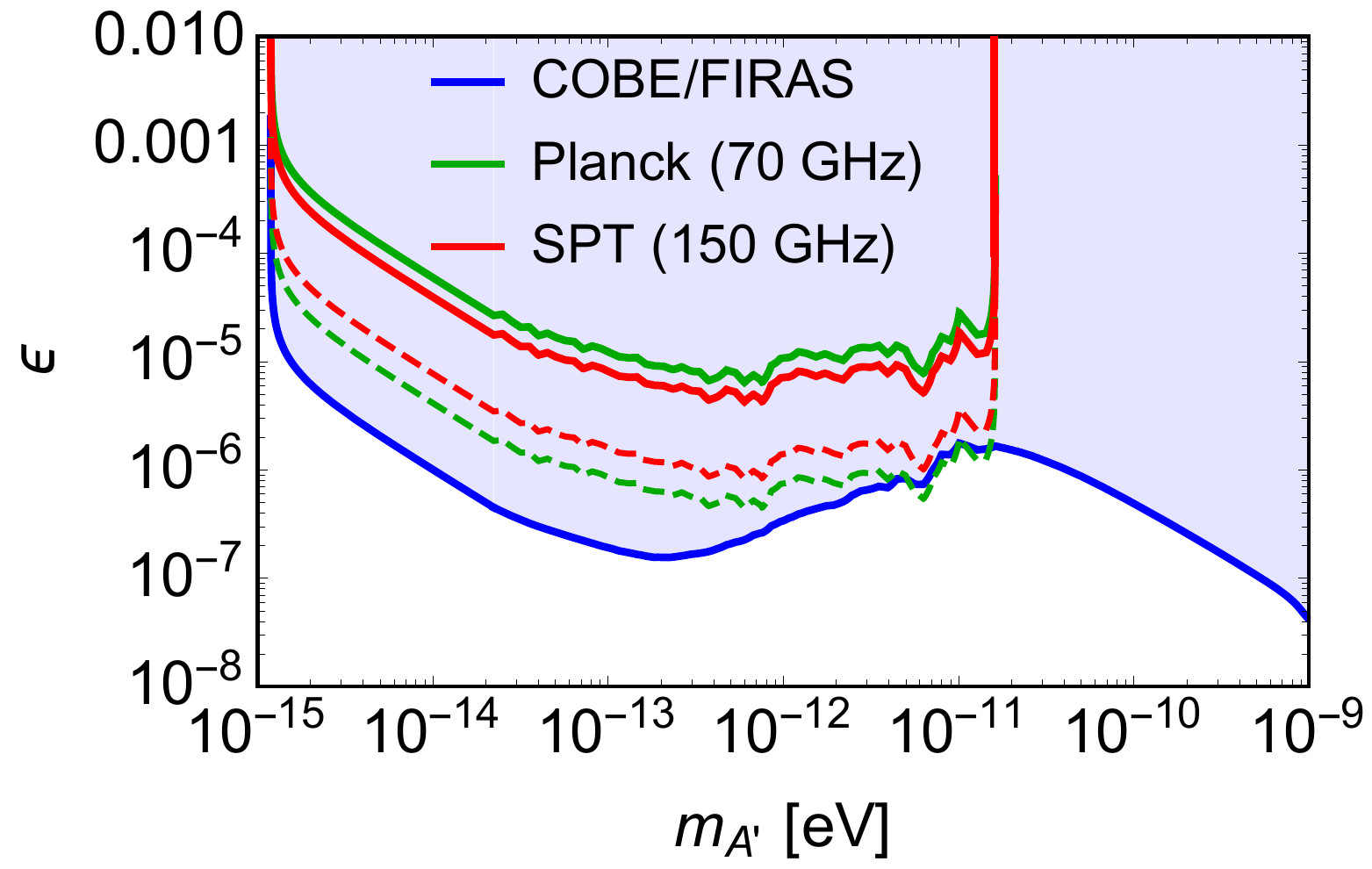}
    \caption{Constraints on the dark photon parameter space from the distortions of the CMB spectrum measured by  COBE/FIRAS  (blue line and shaded region, as in Fig.~\ref{fig:eps}), and estimates of the constraints from CMB anisotropies measured by Planck (green line) and SPT (red line). The solid green and red lines assume that the variance of the converted flux is smaller by a factor $N_{\text{LOS}}$
    mimicking the finite resolution of Planck (SPT), while dashed lines show the exclusion without penalty, see text for details.}
    \label{fig:eps2}
\end{figure}

As alluded to above, a qualitatively different constraint can be obtained from measured CMB anisotropies, that are mapped with a high precision by the Planck satellite~\cite{2018arXiv180706206P} and the SPT~\cite{article_SPT}, among other probes.  
To put a constraint, we use the obtained variance in $P_{\rm tot}$ and 
follow the same procedure that is described in the companion paper~\cite{Bondarenko:2020moh} where anisotropies were used to limit dark photon conversion instead. Concretely, we calculate the standard deviation for the spectral radiance along random continuous LOS,
\begin{equation}
    \delta B_\omega(\omega) = B^{\text{CMB}}_{\omega}  \delta P_{\text{tot}}(\omega)
\end{equation}
and compare with the CMB flux variance measured by all-sky surveys at the exemplary frequencies $\nu = 70$~GHz (Planck) and  $\nu = 150$~GHz (SPT). It is important to note, that the LOS width at a cosmological distance away from $z=0$ is typically smaller than the width that corresponds to the best angular resolution of Planck (SPT).
To make an estimate we assume that LOS within the angular resolution of Planck (SPT) are independent and we divide the variance by a factor $N_{\rm LOS}(z=6)$ that corresponds to the number of continuous LOS that are required to fill ``one Planck (SPT) angular pixel'' at $z=6$, $N_{\rm LOS} \simeq 43000\, (670)$ for Planck (SPT). 

The contours where the CMB flux variance is equal to the variance induced by the loss of CMB photons from resonant conversion is shown in Fig.~\ref{fig:eps2}. Two contours are shown: one for which the conversion variance was divided by the number of LOS in one pixel (solid lines) and one, where the variance of our simulations is left unchanged (dashed lines). We expect that the actual sensitivity lies in between both contours, likely closer to the solid lines. 

There are two possible improvements on the CMB anisotropy constraint. Firstly, in the estimate above we used the all-sky variance of the CMB flux that corresponds to the all-sky temperature anisotropy $\delta T_{\text{CMB}}/T_{\text{CMB}} = \sum_{\ell} (2\ell +1)/(4\pi)C_{\ell} \approx 4\times 10^{-5}$. However, if the signal is most prominent on an angular scale $\theta < \theta_{\max}$, a stronger constraint using Planck (SPT) data can be put using  multipoles $\ell \gtrsim \pi / \theta_{\max}$ in the above sum. 
Secondly, we use a very simple and robust exclusion condition in that the total variance in the flux of converted photons should be smaller than the total CMB anisotropy. One may, however, instead compute the power spectrum of converted flux and hence constrain excess power in individual $\ell$-modes. 
For a more detailed discussion on using CMB anisotropies as a probe for dark photon conversion, see the companion paper~\cite{Bondarenko:2020moh}.

\section{Conversion of dark photon into usual photons}
\label{sec:inverse_conversion}

The phenomenology of resonant conversion becomes richer when considering pre-existing fluxes of dark photons that can be converted into the visible sector. A simple possibility is to source such dark radiation through DM decays. 
For example, a realization that was considered in~\cite{Pospelov:2018kdh} is to use an axion-like particle $a$ as DM. Supplementing (\ref{L}) with 
\begin{equation}
    \mathcal{L}' = \frac{1}{2} (\partial_{\mu} a)^2 - \frac{m_a^2}{2} a^2 + \frac{a}{4 f_a} F'_{\mu \nu} \tilde{F}^{' \mu \nu} 
    \label{eq:Linv}
\end{equation}
induces a two-body decay $a\to A'A'$ that sources dark radiation in form of $A'$ particles. Here, $m_a$ is the DM mass and $f_a$ is the axion decay constant. Note that since the DM decay rate $\Gamma_{a\to A'A'}\propto m_a^3 f_a^{-2}$, sub-eV mass DM requires a low-scale axion with $f_a\lesssim 1\,\rm TeV$ to obtain any appreciable decay rate (that still satisfies  $\Gamma_{a\to A'A'}/H_0 \lesssim 0.1$ in agreement with general bounds on decaying DM~\cite{Poulin:2016nat}). As the link to the Standard Model is protected through the kinetic mixing portal, the stellar cooling constraints only require  $\epsilon/f_a \lesssim 2\times 10^{-9}\,\GeV$~\cite{Pospelov:2018kdh}.

The model above was considered to explain the EDGES anomaly~\cite{Bowman:2018yin}. 
In this work we derive detailed  spectral and spatial properties of this signal without relation to EDGES by making use of dark matter and free electron density maps obtained from the \textsc{eagle} simulation.
The application of these results to EDGES and corresponding constraints are reported in ourr companion paper~\cite{Bondarenko:2020moh}.

A dark photon created from DM 2-body decay  at redshift $z_{\text{dec}}$ will have an initial energy $m_a/2$. Propagating towards Earth, it may then be converted to a normal photon at some redshift $z_{\text{res}}$.
Once the photon, created in this way, arrives at Earth, its frequency $\omega$ will be given by 
\begin{equation}
    \omega = \frac{m_a}{2(1 + z_{\text{dec}})}.
    \label{eq:zdec}
\end{equation}

Photons that are produced by resonances before recombination, $z_{\rm dec} \geq 1700$, may be lost by free-free absorption (inverse Bremsstrahlung), inducing $y$-type spectral distortions instead~\cite{Chluba:2015hma}. The optical depth against absorption is inversely proportional to the cubic power of the photon energy, and the Universe becomes transparent to photons that are born with initial energy in excess of $\sim 1~\rm GHz$ at recombination~\cite{Chluba:2015hma} (hence, in principle, observable at $\rm MHz$ frequencies today.) As we are concerned with higher energy quanta, we may consider the Universe as transparent post recombination. 
 
The observed differential flux of photons at Earth can then be written as
\begin{equation}
    \mathcal{F}_A(\omega) = 
    \mathcal{F}_{A'}^{\text{no conv}}(\omega) P_{\text{tot}}(\omega), 
    \label{eq:flux_def}
\end{equation}
where for brevity we have introduced a notation for the frequency- and angle-differential flux,
$\mathcal{F}(\omega) \equiv {d F}/{d\omega d\Omega}(\omega)$ and generally suppressed the dependence on the LOS direction in all quantities.
The first factor, $\mathcal{F}_{A'}^{\text{no conv}}$, describes the differential flux of dark photons from DM decays in the case if no conversion happened between decay point and Earth, while $P_{\text{tot}}$ is the total probability that a dark photon is converted into a photon while it propagates from $z_{\text{dec}}$ to Earth, given by Eq.~\eqref{eq:ptot} with $z_{\rm max} = z_{\text{dec}}$.

The dark photon flux is given by (a derivation is provided in Appendix~\ref{sec:master-formula}; see also previous works~\cite{Cui:2017ytb,Pospelov:2018kdh})
\begin{align}
    &\mathcal{F}_{A'}^{\text{no conv}} (\omega) = 
    \frac{\Gamma_{a \to A' A'}}{2\pi H(z_{\text{dec}}) m_a} 
   \frac{\rho_{\text{DM}}(z_{\text{dec}})}{\omega(1+z_{\text{dec}})^3} \, \theta\left(\frac{m_a}{2} -\omega \right), 
    \label{eq:flux-dark-photon}
\end{align}
where  $z_{\text{dec}}$ is expressed via the frequency $\omega$ and axion mass $m_a$ through Eq.~\eqref{eq:zdec}, $\rho_{\text{DM}}(z)$ is the DM density along the line of sight and $\theta$ is a unit step function. Note that while in previous contexts this formula has been applied to the cosmological distribution of DM, here it is written in a way that is suitable to be evaluated for the actual DM distribution along a continuous LOS.

The photon flux in Eq.~\eqref{eq:flux_def} is proportional to the DM density at the decay redshift $z_{\text{dec}}$, that is connected to the photon frequency via~\eqref{eq:zdec}.
Therefore, the spectral shape of the signal is proportional to the DM density distribution along the line of sight $\rho_{\text{DM}}(z)$.
The details of $\rho_{\text{DM}}(z)$ affect only the spectrum of the signal, but not the overall number of dark photons produced.

\begin{figure}[t]
    \centering
    \includegraphics[width=0.48\textwidth]{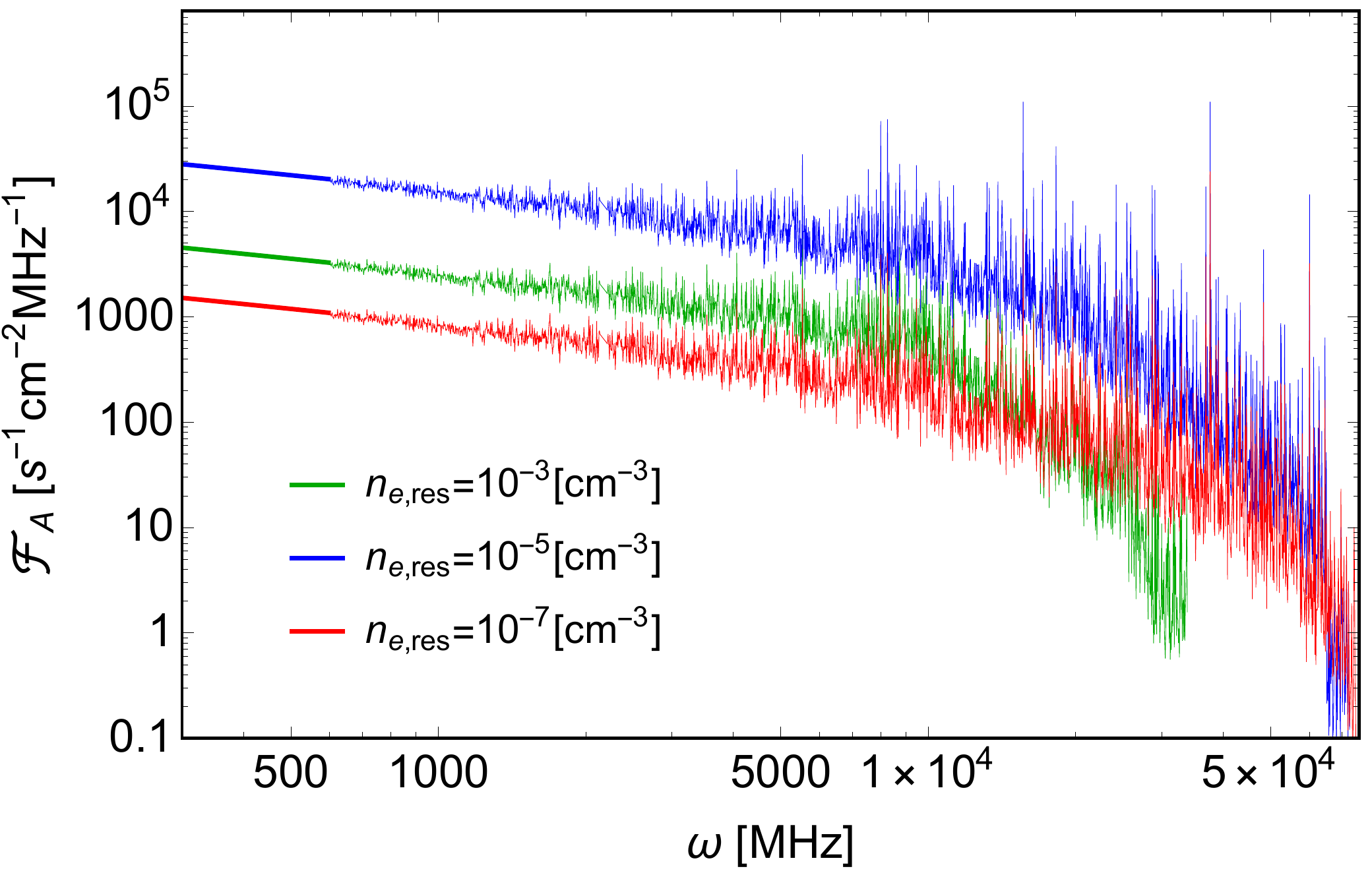}~\includegraphics[width=0.48\textwidth]{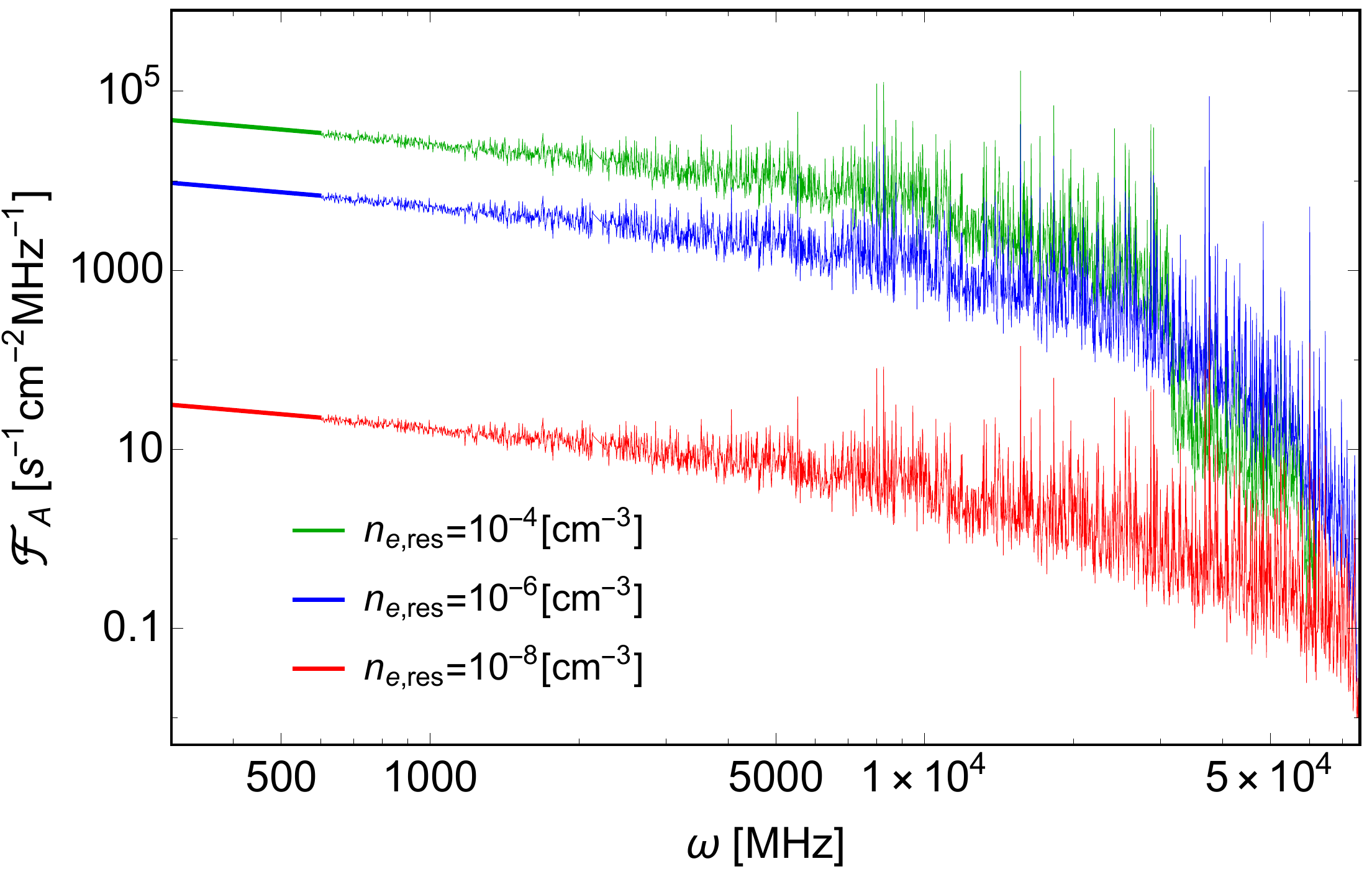}
    \caption{The expected photon flux at Earth  as a function of frequency for different values of resonance number density.  To produce this figure we fixed $m_a=10^{-4}$~eV, $\epsilon=10^{-9}$, $\Gamma_{a\to A' A'} = (10 ~t_{0})^{-1}$, where $t_{0}= 13.8\,{\rm Gyr}$ is the age of the Universe.}
    \label{fig:flux}
\end{figure}

We now use our simulation to obtain properties of the signal in this model.
Figure~\ref{fig:flux} shows examples of converted photon flux as a function of frequency for 6 values of resonance number densities (dark photon mass). 
To elucidate the dependence on the model parameters, let us introduce the dimensionless variable $x = 2\omega/m_a$. The photon flux at Earth~\eqref{eq:flux_def} can be rewritten as
\begin{equation}
    \mathcal{F}_{A}(\omega) = 2 \epsilon^2 \Gamma_{a\to A'A'}\frac{m_{A'}^2}{m_a^3} f\left(\frac{2\omega}{m_a}\right),
    \qquad
    f(x) = x \frac{\rho_{\text{DM}}[z_{\text{dec}}(x)]}{H[z_{\text{dec}}(x)]} \sum_i \frac{R_i}{1+z_{i}}.
    \label{eq:flux-through-x}
\end{equation}
The dependence on $m_a$ enters through the ratio $x = 2\omega/m_a$ and, additionally, $m_a$ defines the overall amplitude $\propto m_a^{-3}$.
The dependence on the frequency is more involved.
Let $z_{\text{res,low}}^{\text{max}}$ be the redshift of the last resonance in the low-redshift region $z<6$ and $z_{\text{res,high}}$ the redshift of the resonance in the high-redshift region $z>20$.
Consider then a region where there are no resonances (for which $z_{\text{dec}}$ is between $z_{\text{res,low}}^{\text{max}}$ and $z_{\text{res,high}}$ or above $z_{\text{res,high}}$). In this region the frequency dependence in Eq.~\eqref{eq:flux-through-x} is defined only by the factor $x \rho_{\text{DM}}[z_{\text{dec}}(x)]/H[z_{\text{dec}}(x)]$. Here $\rho_{\text{DM}} \propto (1+z_{\text{dec}})^3 \propto x^{-3}$, while $H \propto (1+z_{\text{dec}})^{3/2}\propto x^{-3/2}$ in the matter dominated epoch  and $H \propto (1+z_{\text{dec}})^{2}\propto x^{-2}$ in the radiation dominated epoch. Therefore,
\begin{equation}
    \mathcal{F}_{A}(\omega) \propto \left\{
    \begin{matrix}
    1/\sqrt{\omega},\text{\quad} & z_{\text{res,low}}^{\text{max}} <z_{\text{dec}}<z_{\text{res,high}}\text{ or } z_{\text{res,high}}< z_{\text{dec}} \ll z_{\text{eq}}, \\
    \text{const},\quad & z_{\text{dec}} \gg z_{\text{eq}},
    \end{matrix}
    \right.
\end{equation}
where $z_{\rm eq} $ is the redshift of matter-radiation equality. For larger frequencies for which $z_{\text{dec}} < z_{\text{res,low}}^{\text{max}}$ decays happen inside the region of low-redshift resonances, less resonances contribute to the conversion and the spectrum is steeper than $\propto 1/\sqrt{\omega}$.

Another effect that we observe in Fig.~\ref{fig:flux} is that for large resonant electron number densities the signal cuts off before the frequency reaches its maximal allowed value $m_a/2$. This happens because in our random LOS approach, the electron number density does not reach the required $n_{e,\text{res}}$ value at low enough redshift, see Fig.~\ref{fig:neDMLOS}. Such conversion is contingent on the LOS passing through the centers of galaxy clusters, which is associated with a small probability as discussed above.

\begin{figure}[t]
    \centering
    \includegraphics[width=0.8\textwidth]{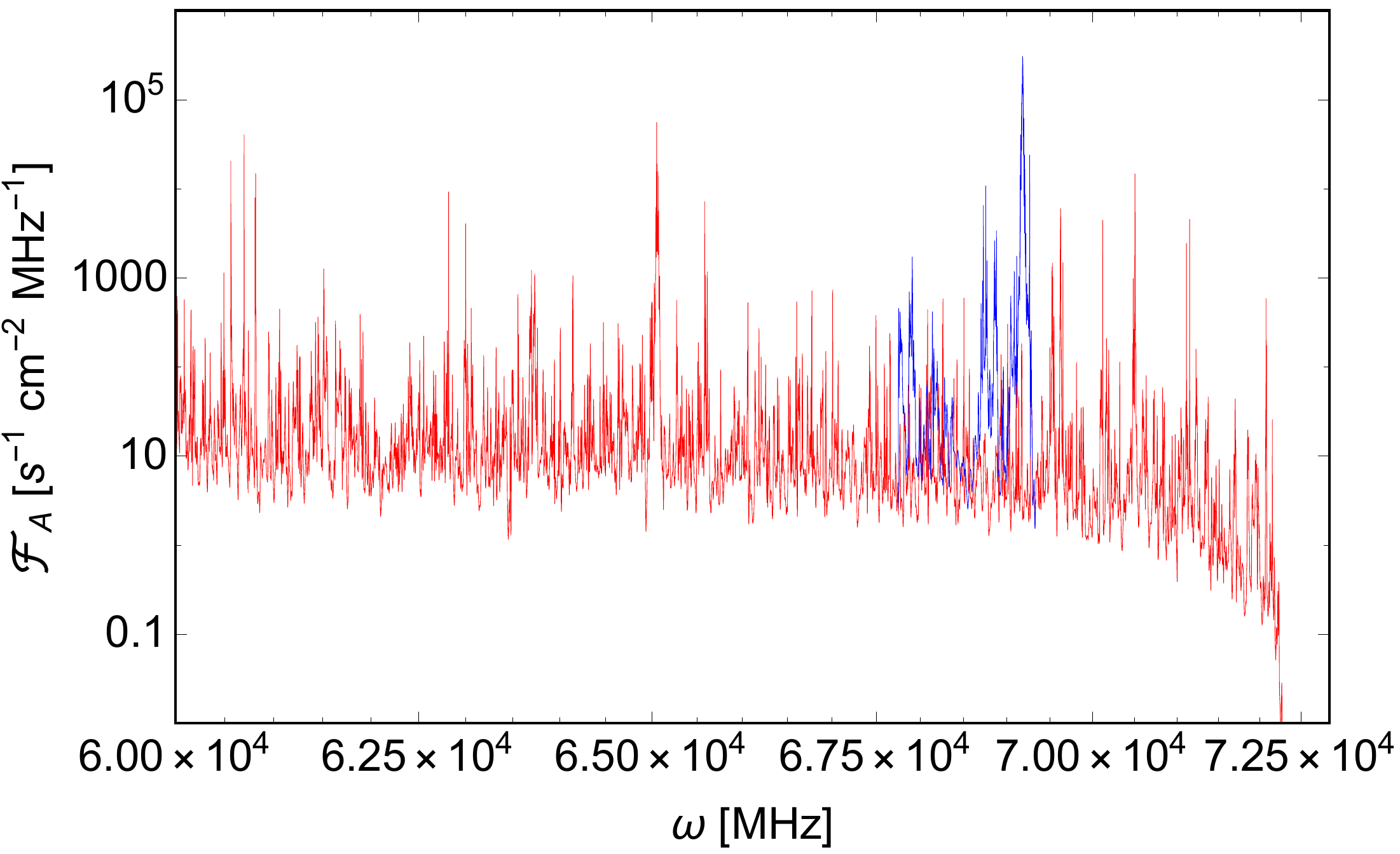}
    \caption{
    An example of photon flux generated along an exemplary LOS (red line)  for the resonant electron number density $n_{e,\text{res}} = 10^{-5}\text{ cm}^{-3}$. The blue line shows the flux where one LOS segment has been replaced in order to cross through a cluster at $z=0.1$. The parameters of the model are chosen as in Fig.~\ref{fig:flux}.
    }
    \label{fig:flux_cluster}
\end{figure}

For concreteness, we shall consider the example of a rare LOS that passes through the center of the most massive cluster in our simulation at $z=0.1$. Given that the signal is proportional to DM density along continuous LOS, one expects a significant increase in flux from DM decays inside the cluster. This translates to a peak at a characteristic frequency, shown by the blue line in  Fig.~\ref{fig:flux_cluster}. Such a signal can be searched for by telescopes with high  energy resolution.  Nevertheless, the encounter of a cluster does not change much the overall conversion probability of the signal from higher redshifts.

Finally, along  different spatial directions on the sky the  flux of converted photons at a given frequency will fluctuate because of variations in the DM density and in the probability of conversion. For a given direction on the sky, the flux, normalized to its mean value is given by
\begin{equation}
    \frac{\mathcal{F}_A(\omega)}{\langle \mathcal{F}_A(\omega) \rangle} =
    \frac{\rho_{\text{DM}}(z_{\text{dec}})}{\langle \rho_{\text{DM}}(z_{\text{dec}}) \rangle} \times
    \frac{P_{\text{tot}}}{\left\langle P_{\text{tot}} \right\rangle},
    \label{eq:FtoFav}
\end{equation}
where $\langle\dots\rangle$ is the all-sky average for fixed $z_{\text{dec}}$ and $n_{e,\text{res}}$. On the RHS we used that DM density and the probability of conversion are statistically independent, as DM decay and dark photon conversion tends to happen at spatially well separated locations. We hence can investigate the  fluctuations in both factors independently.

\begin{figure}[t]
    \centering
    \includegraphics[width=0.48\textwidth]{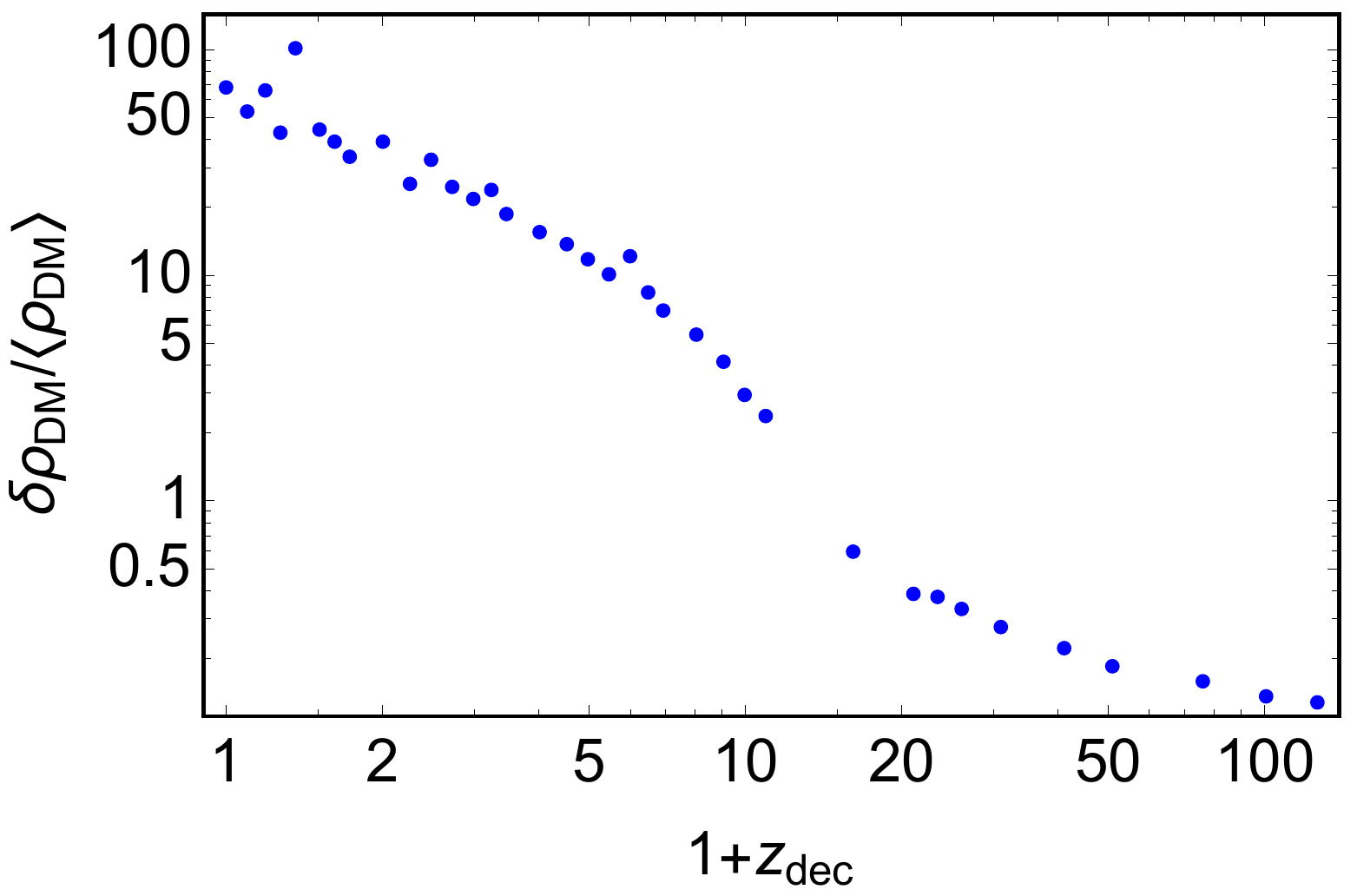}
    \caption{The standard deviation of the DM density normalized to the average DM density as a function of decay redshift. To produce this figure we generated 1000 random LOS for DM density in each snapshot.
    }
    \label{fig:deltaRconvRDM}
\end{figure}

The variance of the second factor in Eq.~\eqref{eq:FtoFav} we have already investigated, see Fig.~\ref{fig:deltaRconv}. To calculate fluctuations originating from the spatial distribution of DM, we used $1000\times 5000$ points at each redshift, corresponding to 1000 LOS.
The result is shown in Fig.~\ref{fig:deltaRconvRDM}. We see that fluctuations of the DM density decrease with redshift as the Universe becomes more homogeneous.
The variances of both factors in Eq.~\eqref{eq:FtoFav} are used in our companion paper~\cite{Bondarenko:2020moh} to constrain the model proposed in~\cite{Pospelov:2018kdh} that is able to resolve the EDGES anomaly.

\section{Conclusions and discussion}

In this paper we study the cosmological distribution of the value of effective photon mass and its effect on  the propagation of light through the Universe. Using \textsc{eagle} simulations, we extract the electron number density along random LOS for the low-redshift, post-reionization Universe at $z<6$. It is tied to the number density of baryons and the growth of non-linear structure make it a highly fluctuating quantity. We find, that a given value of $n_e$, within a broad interval $10^{-9} \lesssim n_e/{\rm cm}^{-3} \lesssim 10^{-1}$, is met a great number of redshifts when intersecting voids and filaments. In contrast, collapsed structures like galaxy clusters are rarely met in a random LOS with a cross-sectional area of $(25\times 20)\,{\rm ckpc}^2$. For higher redshifts, $20 \leq z \leq 125$, we use DM density as a proxy to infer the value of $n_e$. 
In this work, we stay clear from the epoch of reionization at $6\lesssim z \lesssim 20$, as the detailed nature of its progression is uncertain. We note, however, that we expect the electron density to be fluctuating significantly at the process of transitioning from a largely charge neutral to ionized Universe proceeds by the growth of patches. More advanced simulations which take into account radiative transfer, such as Aurora~\cite{2017MNRAS.466..960P} are needed in this case. We leave a study of latter point for future work. 

The effective mass of the photon can be important for  models where photons kinetically  mix with a dark vector particle, generally referred to as dark photon. The value of the effective photon mass after recombination is in the the range $10^{-15} \lesssim m_{A}/\eV \lesssim 10^{-9}$, and the mere presence of a dark photon with a mass in that interval allows for the resonant conversion between both states. 

In a first part, we study generic constraints on the $\epsilon$-$m_{A'}$ parameter space that are independent on any cosmological population of dark photons. Using the CMB as a precision probe, we study the loss CMB radiation quanta during propagation and investigate additional anisotropies that are being imprinted in this process. Using the spectral measurements of the COBE/FIRAS instrument, values of $\epsilon$ in the ballpark of $10^{-6}$ are constrained by  deviations from the blackbody law, Fig.~\ref{fig:eps}. Resolving the inhomogeneities in the late time Universe, allows us to improve previous constraints~\cite{Mirizzi:2009iz, Kunze:2015noa} on three decades in dark photon mass, and in good agreement with independent similar recent work~\cite{Caputo:2020bdy}.

In Figure~\ref{fig:eps2} we then present the estimate of the constraint that can be derived from conversion-induced excess temperature anisotropy in the {\em all-sky} observed variance in the 70~GHz and 150~GHz channels of Planck and SPT, respectively. This is the first constraint of this sort on the dark photon model, and kinetic mixing angles $\epsilon \gtrsim 10^{-4}$ are conservatively excluded in the range $10^{-15} \lesssim m_{A'}/\eV \lesssim 10^{-11}$. The latter constraint can be (significantly) strengthened by computing the angular power-spectrum of the modified signal and comparing it to the observed values of $C_\ell$, mode by mode in~$\ell$; we leave such investigation for future work.

In a second part, we consider the case of a  dark radiation component in form of dark photons. For concreteness, we assume that $A'$ were sourced monochromatically, in the 2-body decay of DM with cosmologically long lifetime, $\tau_a \gg H_0^{-1}$. The great number of resonances together with the instantaneous conversion into ordinary photons, produce  observable fluxes that carry tomographic imprints of the intervening electron density along any random LOS, Fig.~\ref{fig:flux}. The overall flux amplitude depends on~$\epsilon^2 / \tau_a$.  
Depending on the value of this combination, the signal from late-time conversion can be constrained both from the measurements of the value of CMB temperature as a function of frequency by COBE/FIRAS and from CMB anisotropies by Planck, SPT, among others. These constraints are explored in the 
 companion paper~\cite{Bondarenko:2020moh}. Finally, we note that the predicted signal can be also probed by  lower-frequency radio telescopes like LOFAR~\cite{2013A&A...556A...2V} or  SKA~\cite{Bacon:2018dui}.
 
Probing larger values of dark photon mass through cosmological  resonant conversion in the late time Universe hinges on the values of maximal encountered electron density. In our  simulations we resolve regions with $\lesssim 0.1\text{ cm}^{-3}$. However, these high values are present along LOS that we consider  only at $z=4-6$, where the average density is high, so these values of $n_e$ do not correspond large overdensities. At lower redshifts,  regions of high electron densities
are associated with the central regions of galaxy clusters and within galaxies.
This will result in an additional contributions to the expected signals from many point-like sources, and constitutes another avenue for further investigation.

The simulated data on $n_e(z)$ and $\rho_{\text{DM}}(z)$ as well as the Mathematica notebook that is able to access this data is made publicly available at the Zenodo platform~\cite{Zenodo}. We provide  functions for the calculation of conversion probability along  continuous LOS and for the signal-generation in the explored model with DM.

\acknowledgments
We would like to thank Alexey Boyarsky for support and collaboration.
We thank Matthieu Schaller for re-running a dark matter only version of the \textsc{eagle} simulation with additional outputs at $z>20$. We acknowledge the Virgo Consortium for making their simulation data available. The \textsc{eagle} simulations
were performed using the DiRAC-2 facility at Durham, managed by the ICC, and the PRACE facility
Curie based in France at TGCC, CEA, Bruyères-le-Châtel. This research was supported in part by the National Science Foundation under Grant No.~NSF PHY-1748958. AS and JP are supported by the New Frontiers program of the Austrian Academy of Sciences. KB is supported by the European Research Council (ERC) Advanced Grant ``NuBSM'' (694896).

\appendix

\section{Effective photon mass in the medium}
\label{sec:effective-mass}

\begin{figure}[t]
    \centering
    \includegraphics[width=0.48\textwidth]{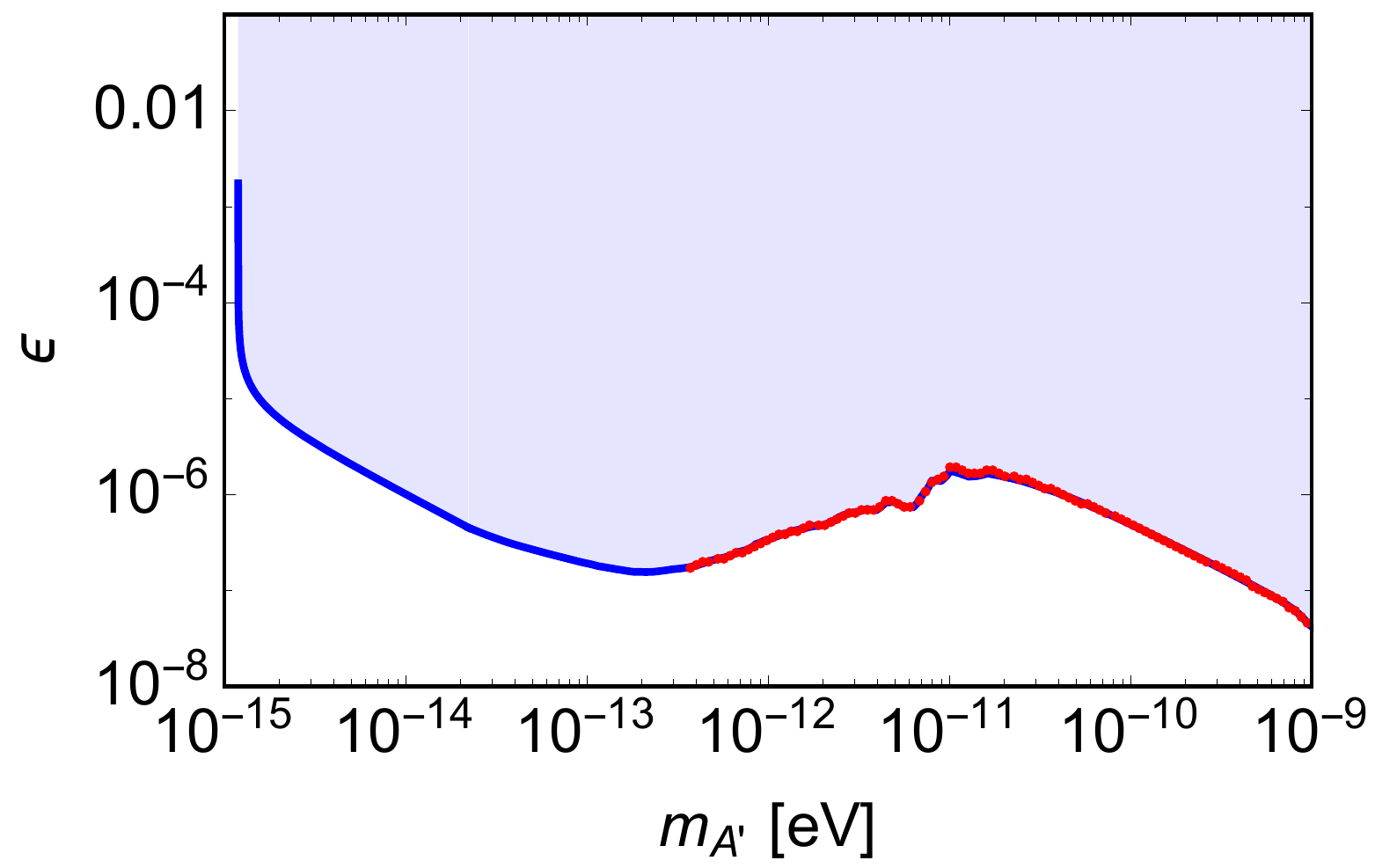}~\includegraphics[width=0.48\textwidth]{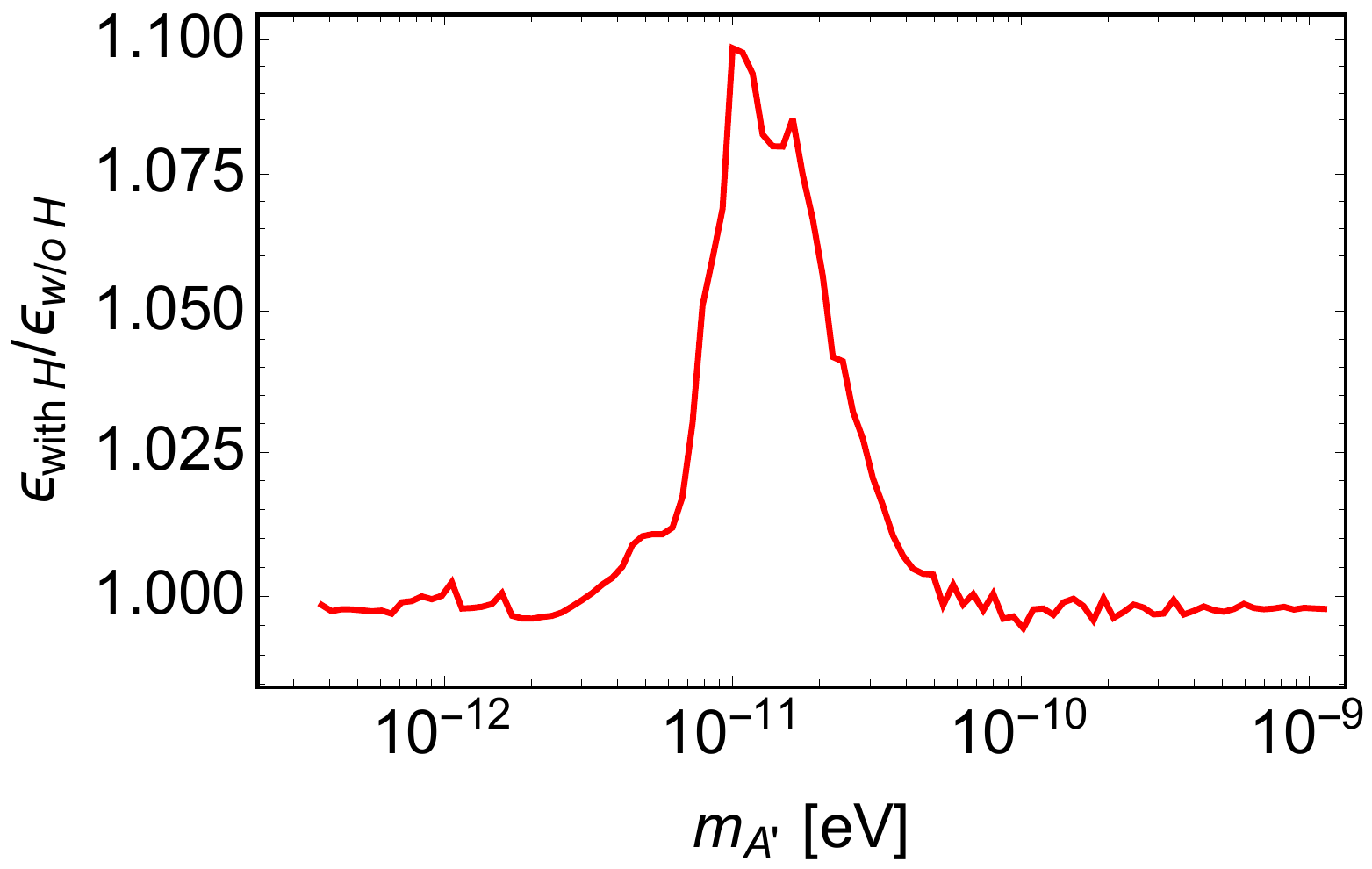}
    \caption{\textit{Left panel}: comparison of the constraint from FIRAS discussed in Section~\ref{sec:Conversion-CMB} without effects from the neutral hydrogen (blue line) and with neutral hydrogen (red line). \textit{Right panel}: the ratio between the constraints on $\epsilon$ shown on the left panel.}
    \label{fig:FIRAS-neutral-H}
\end{figure}

 In this work, we are interested in the conversion between physical one-particle states, hence only  transverse polarizations are relevant. The general dispersion relation of a photon with energy $\omega$ and three momentum $\vec k$ is the solution to $\omega^2 - \vec k^2 - \real \Pi_{\rm T,L}(\omega, \vec k) = 0$ where  $\Pi_{\rm T}(\omega, \vec k)$ is the in-medium polarization function, see e.g.~\cite{Braaten:1993jw}. Notwithstanding a generally complicated dependence of frequency $\omega(\vec k)$ and ``effective photon mass'' on photon momentum, away from resonances of bound electrons and in an isotropic, non-degenerate and non-relativistic medium and to leading order in $\alpha$ the latter takes a constant form, $ \real \Pi_{\rm T,L} \simeq \omega_p^2$ where $ \omega_p^2 \simeq 4\pi \alpha n_e/m_e$  is the squared plasma frequency. Hence, we identify $m_A^2 = \omega_p^2$ from Eq.~(\ref{eq:mA}). For sub-eV frequencies, when including the contribution from neutral hydrogen, the expression is modified to~\cite{Mirizzi:2009iz}
\begin{equation}
   \label{mA}
     m_{A}^2 \simeq \omega_p^2 - 2 \omega^2 (n-1)|_{\rm H}. 
\end{equation}
Here,   $(n-1)|_{\rm H} \simeq 13.6\times 10^{-5}$ ($\omega \leq 1\,\eV$) is the neutral hydrogen gas index of refraction~\cite{Peck:77}. 
Corrections to (\ref{mA}) arise from the finite temperature of the gas, from ions, as well as from neutral helium. All these contributions are suppressed and can be neglected for the purpose of this paper.
Figure~\ref{fig:FIRAS-neutral-H} explores the influence of the neutral hydrogen component on the resulting constraints from COBE/FIRAS. As can be seen, results are practically identical with at most a 10\%\ correction, and we may take $m_A^2 = \omega_p^2$, justifying Eq.~(\ref{eq:mA}).

\section{Dependence on LOS averaging width}
\label{sec:thin-thick-LOS}

\begin{figure}[t]
    \centering
    \includegraphics[width=0.48\textwidth]{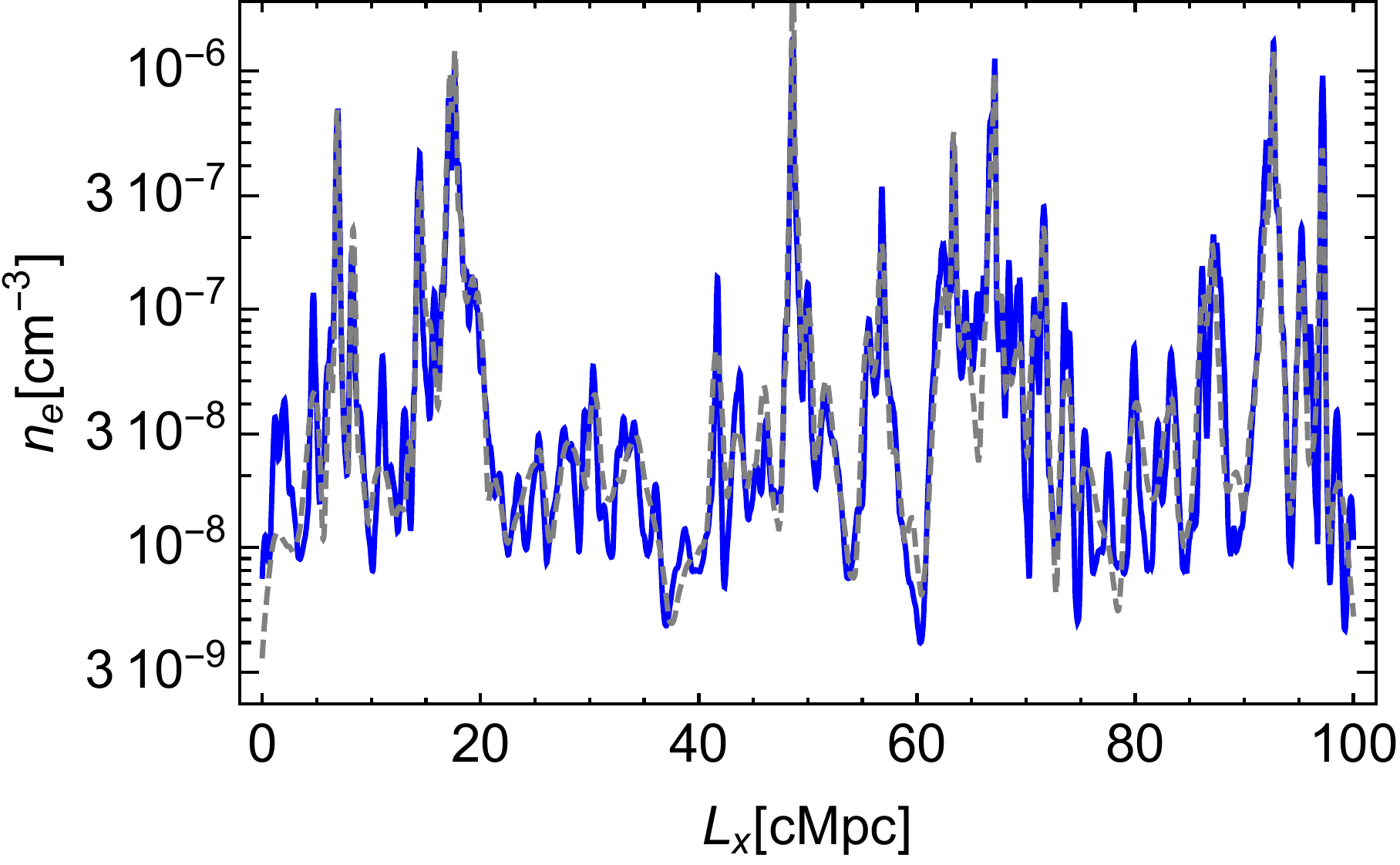}~\includegraphics[width=0.48\textwidth]{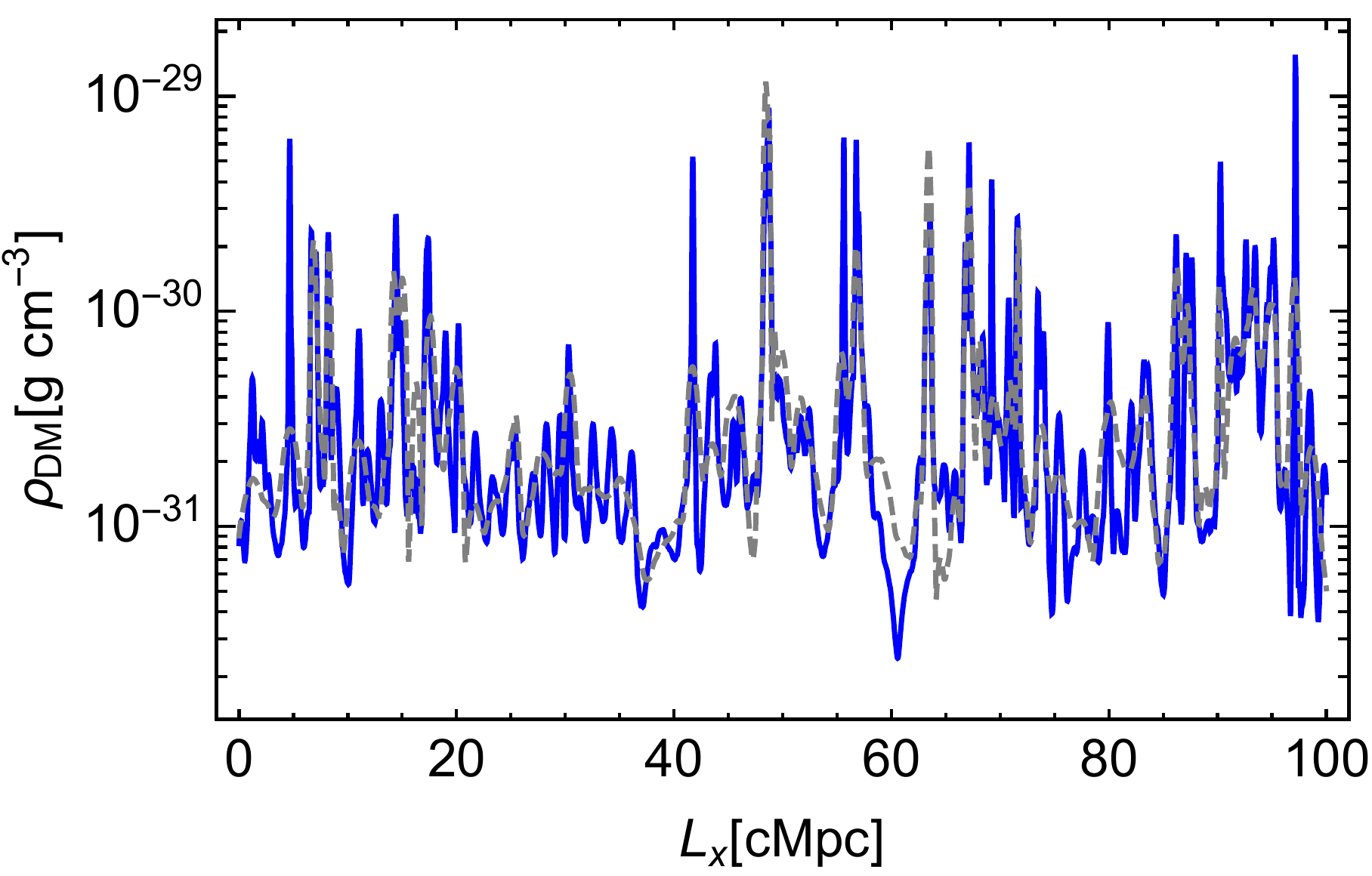}
    \caption{An example of the electron number density (left panel) and DM density (right panel) for the same LOS at $z=0$, calculated using a thicker $250$~ckpc (gray dashed line) and thiner $25$~ckpc (blue line) averaging width. $L_x$ is a distance along the LOS within the simulation box.}
    \label{fig:slice_25_250}
\end{figure}

In this work we use LOS that are extracted from simulation slices of constant comoving width of $25$~ckpc. In our companion paper~\cite{Bondarenko:2020moh} we use thicker slices of $250$~ckpc width. Both yield identical results on the average conversion probability within their standard errors. 

\begin{figure}[t]
    \centering
    \includegraphics[width=0.48\textwidth]{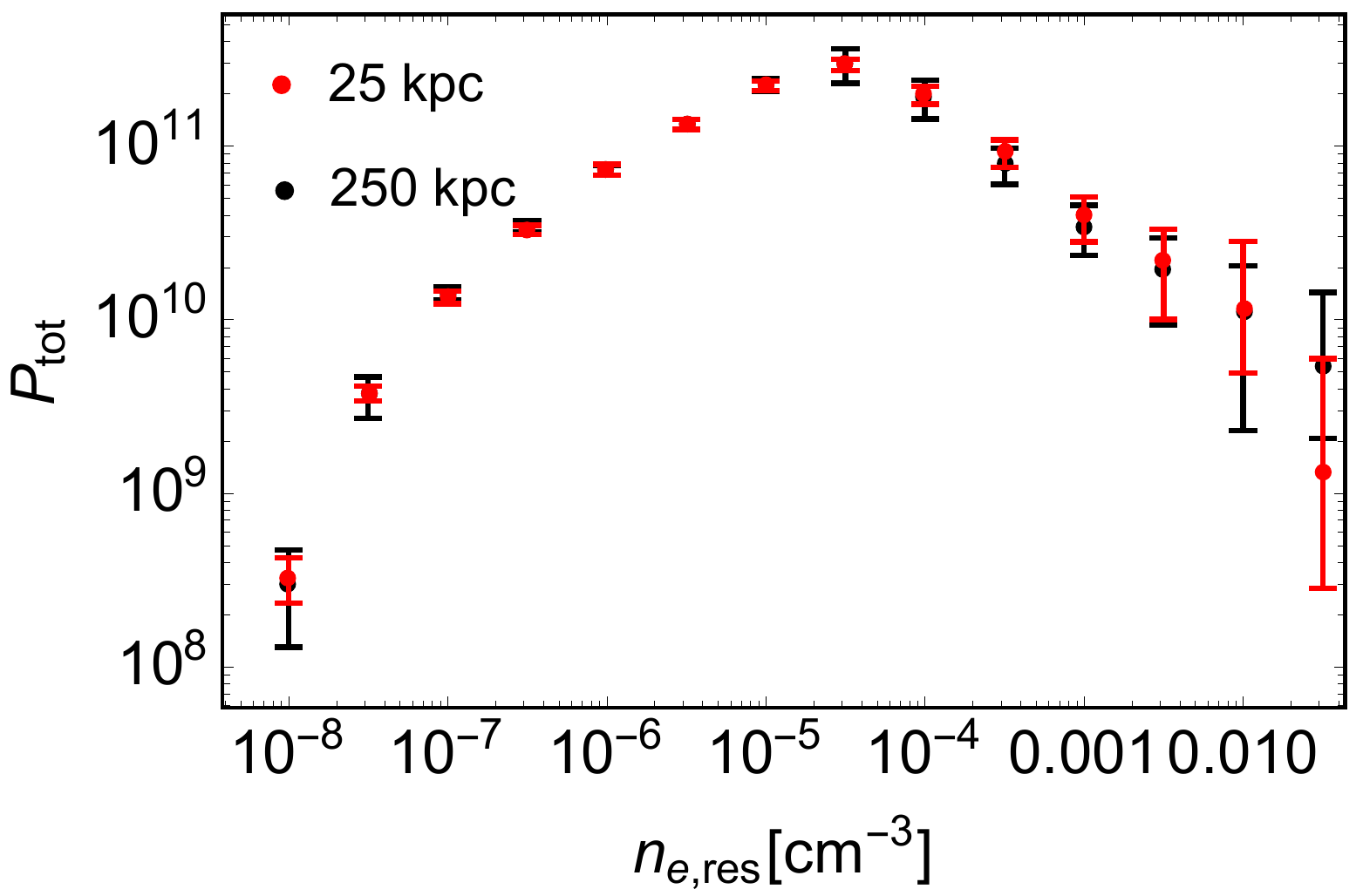}
    \caption{Comparison of the total conversion probability calculated over 100 continuous LOS from two samples of simulations with thin ($25$~ckpc, red dots) and thick ($250$~ckpc, black dots) LOS. The error bars show the standard deviation.}
    \label{fig:Ptot_25_250}
\end{figure}

Figure~\ref{fig:slice_25_250} shows the electron number density and the DM density along the same LOS at redshift $z=0$.  Although the thinner LOS contains more small-scale fluctuations than the thicker one, it has practically no effect on the overall conversion probablity, as demonstrated in Fig.~\ref{fig:Ptot_25_250}. This is concordant with the observation that even in presence of a large number of resonance crossings, $P_{\rm tot}$ is at most a factor of unity different from the probability inferred from cosmologically averaged densities, see Fig.~\ref{fig:PtotInhomoToHomo}.

\section{Photon flux from dark photon conversion}
\label{sec:master-formula}

For completeness, here we present a derivation of the photon flux from  conversion of dark photons that were created in DM decays. 
Consider a telescope with effective area $dA$ and angular resolution $d\Omega \ll 1$ that is located at a redshift of observation $z_{\text{obs}}$.
The photon flux depends on the number density of decaying DM, $n_{\text{DM}}(\ell)$, its mass $m_a$, its decay rate $\Gamma_{a\to A' A'}$, on the electron number density $n_e(\ell)$ and on the kinetic mixing strength~$\epsilon$, where $\ell$ is a distance between the telescope and the given point along the line of sight at the time of observation.

Given the negligible kinetic energy of DM with respect to its rest mass (a ratio of $10^{-4} - 10^{-6}$), $A'$ are always created with the same initial energy $m_a/2$ in the assumed 2-body decay. The energy during conversion does not change, and assuming relativistic particles throughout, neither does their momentum. The frequency of converted photons $\omega$ measured by the telescope depends on the decay redshift $z_{\text{dec}}$ at which $A'$ was injected,
\begin{equation}
    \omega = \frac{m_a (1 + z_{\text{obs}})}{2(1 + z_{\text{dec}})}.
\end{equation}

The flux of photons near the telescope is equal to the flux of dark photons near the telescope without conversion, times the total probability of dark photon conversion along its LOS until it reaches the observer 
\begin{equation}
    \frac{d F_{A}}{ d\omega d\Omega}(\omega) =
    \frac{d F_{A'}^{\text{no conversion}}}{d\omega d\Omega}(\omega) P_{\text{tot}}(\omega).
    \label{eq:flux_def2}
\end{equation}

Consider a DM region with a number density $n_{\text{DM}}(z_{\text{dec}})$ that is located in the infinitesimal redshift interval $z_{\text{dec}}$ and $z_{\text{dec}}+dz_{\text{dec}}$. Let us find the flux of dark photons near the telescope considering that there is no conversion.
The number of DM particles in this region is
\begin{equation}
    dN_{\text{DM}}(z_{\text{dec}}) = n_{\text{DM}}(z_{\text{dec}}) dV = n_{\text{DM}}(z_{\text{dec}}) \ell_{\text{dec}}^2 d\ell_{\text{dec}} d\Omega, 
\end{equation}
where $\ell_{\text{dec}}$ is the distance between the telescope and the DM region at the time of decay, $\ell_{\text{dec}} = \ell (1+z_{\text{obs}}) / (1+z_{\text{dec}})$, so
\begin{equation}
    dN_{\text{DM}}(z_{\text{dec}}) = \frac{(1+z_{\text{obs}})^3 n_{\text{DM}}(z_{\text{dec}})}{(1+z_{\text{dec}})^3} \ell^2 d\ell d\Omega.
\end{equation}

The number of dark photons produced per time and energy is
\begin{align}
    \frac{dN_{A'}}{dt dE}(z_{\text{dec}}) &= 2 \Gamma_{a \to A' A'} dN_{\text{DM}} \delta\left(E(z_{\text{dec}}) - \frac{m_a}{2}\right).
\end{align}

The number of dark photons that were emitted in the direction of the telescope with area $dA$ is 
\begin{align}
    \frac{dN_{A'\to \text{telescope}}}{dt dE}(z_{\text{dec}}) &=  \frac{dA}{4\pi \ell^2} \frac{dN_{A'}}{dt dE}(z_{\text{dec}}) = \\ &=
    \frac{\Gamma_{a \to A' A'} (1+z_{\text{obs}})^3 n_{\text{DM}}(z_{\text{dec}})}{2\pi (1+z_{\text{dec}})^3} \delta\left(E(z_{\text{dec}}) - \frac{m_a}{2}\right) d\ell d\Omega dA.
    \label{eq:dN-dt1}
\end{align}

The frequency of emitted dark photons at arrival reads,
\begin{equation}
    \omega(z_{\text{obs}}) = \frac{1+z_{\text{obs}}}{1+z_{\text{dec}}}
    E(z_{\text{dec}}).
\end{equation}
The time interval $dt$ between photons increases as 
\begin{equation}
    dt(z_{\text{obs}}) = \frac{1+z_{\text{dec}}}{1+z_{\text{obs}}} dt(z_{\text{dec}}).
\end{equation}

Changing variables, the flux of dark photons at arrival reads
\begin{align}
    \frac{dN_{A'\to \text{telescope}}}{dt d\omega d\Omega dA} &=
    \frac{\Gamma_{a \to A' A'} (1+z_{\text{obs}})^4 n_{\text{DM}}(z)}{2\pi (1+z_{\text{dec}})^3} \frac{1}{\omega} \delta\left(z_{\text{dec}} - \frac{m_a (1+z_{\text{obs}})}{2\omega} + 1\right) d\ell.
\end{align}

The physical distance along LOS to the point of DM decay (given by $z_{\text{dec}}$), at the time of observation, is given by
\begin{equation}
    \ell = \frac{1}{1+z_{\text{obs}}} \int\limits_{z_{\text{obs}}}^{z_{\text{dec}}} \frac{dz}{H(z)},
\end{equation}
where $H(z)$ is a Hubble rate. Using this relation, we find the total flux of dark photons at arrival at the telescope, $F_{A'}^{\text{no conversion}} = dN_{A'\to \text{telescope}}/(dtdA)$, from all regions along the line of sight in its energy and angular-differential form
\begin{align}
    \frac{d F_{A'}^{\text{no conversion}}}{d\omega d\Omega} (\omega) &= 
    \int\limits_{0}^{\infty} \frac{\Gamma_{a \to A' A'} (1+z_{\text{obs}})^4 n_{\text{DM}}(z_{\text{dec}})}{2\pi (1+z_{\text{dec}})^3} \frac{1}{\omega}\delta\left(z_{\text{dec}} - \frac{m_a (1+z_{\text{obs}})}{2\omega} + 1\right) d\ell
     \\ &=
    \frac{\Gamma_{a \to A' A'}(1+z_{\text{obs}})^3}{2\pi H(z_{\text{dec}}[\omega,z_{\text{obs}}])}
    \frac{n_{\text{DM}}(z_{\text{dec}}[\omega,z_{\text{obs}}])}{(1+z_{\text{dec}}[\omega,z_{\text{obs}}])^3}
    \frac{\theta\left(\frac{m_a}{2}- \omega\right)}{\omega}, 
    \label{eq:flux-dark-photon2}
\end{align}
where $\theta(x)$ is a Heaviside step function and
\begin{equation}
    z_{\text{dec}}[\omega,z_{\text{obs}}] = \frac{m_a (1 + z_{\text{obs}})}{2 \omega} - 1.
\end{equation}

At a resonance at redshift $z_{\rm res}$, a dark photon of energy $\omega$ converts to an ordinary photon with probability
\begin{equation}
    P_{A'\to A} = \frac{\pi \epsilon^2 m_{A'}^2}{\omega} R,\qquad R = \left|\frac{d \log n_e}{d \ell}\right|^{-1}_{z_{\text{res}}},
    \label{eq:prob2}
\end{equation}

Let us find the total conversion probability for dark photons of frequency $\omega$ at arrival. For this, assume that the probability in each conversion is much smaller than unity and the total conversion probability $P_{\text{tot}}(\omega) \lesssim 1$. Then we can ignore back-conversions and sum up the individual probabilities along the LOS,
\begin{align}
    P_{\text{tot}}(\omega) =
    \frac{\pi \epsilon^2 m_{A'}^2}{\omega} \sum_i \frac{R_i}{(1+z_{i})} \theta(z_{\rm dec} - z_i).
    \label{eq:Ptot}
\end{align}

Finally, substituting~\eqref{eq:flux-dark-photon2} and~\eqref{eq:Ptot} in~\eqref{eq:flux_def2} we arrive at the formula for the photon flux near the telescope, in agreement with previous expressions~\cite{Pospelov:2018kdh},
\begin{equation}
    \frac{d F_{A}}{d\omega d\Omega}(\omega) =  \frac{\Gamma_{a \to A' A'}}{2 \pi \omega H(z_{\text{dec}}[\omega,z_{\text{obs}}])}
    \frac{(1+z_{\text{obs}})^3 n_{\text{DM}}(z_{\text{dec}}[\omega,z_{\text{obs}}])}{(1+z_{\text{dec}}[\omega,z_{\text{obs}}])^3} P_{\text{tot}}(\omega).
    \label{eq:master-formula}
\end{equation}

\bibliographystyle{JHEP}
\bibliography{ship.bib}

\providecommand{\href}[2]{#2}\begingroup\raggedright\begin{thebibliography}{10}

\bibitem{BornWolf}
M.~Born and E.~Wolf, {\em Principles of Optics: Electromagnetic Theory of
  Propagation, Interference and Diffraction of Light}.
\newblock Cambridge University Press, Cambridge, 1999.

\bibitem{2010MNRAS.408.2163A}
M.~A. {Arag{\'o}n-Calvo}, R.~{van de Weygaert}, and B.~J.~T. {Jones}, {\it
  {Multiscale phenomenology of the cosmic web}},  {\em \mnras} {\bf 408} (Nov.,
  2010) 2163--2187, [\href{http://arxiv.org/abs/1007.0742}{{\tt
  arXiv:1007.0742}}].

\bibitem{Debackere2020}
S.~N.~B. {Debackere}, J.~{Schaye}, and H.~{Hoekstra}, {\it {The impact of the
  observed baryon distribution in haloes on the total matter power spectrum}},
  {\em \mnras} {\bf 492} (Feb., 2020) 2285--2307,
  [\href{http://arxiv.org/abs/1908.05765}{{\tt arXiv:1908.05765}}].

\bibitem{Schaye2015}
J.~{Schaye}, R.~A. {Crain}, R.~G. {Bower}, M.~{Furlong}, M.~{Schaller},
  T.~{Theuns}, C.~{Dalla Vecchia}, C.~S. {Frenk}, I.~G. {McCarthy}, J.~C.
  {Helly}, A.~{Jenkins}, Y.~M. {Rosas-Guevara}, S.~D.~M. {White}, M.~{Baes},
  C.~M. {Booth}, P.~{Camps}, J.~F. {Navarro}, Y.~{Qu}, A.~{Rahmati},
  T.~{Sawala}, P.~A. {Thomas}, and J.~{Trayford}, {\it {The EAGLE project:
  simulating the evolution and assembly of galaxies and their environments}},
  {\em \mnras} {\bf 446} (Jan, 2015) 521--554,
  [\href{http://arxiv.org/abs/1407.7040}{{\tt arXiv:1407.7040}}].

\bibitem{Crain2015}
R.~A. {Crain}, J.~{Schaye}, R.~G. {Bower}, M.~{Furlong}, M.~{Schaller},
  T.~{Theuns}, C.~{Dalla Vecchia}, C.~S. {Frenk}, I.~G. {McCarthy}, J.~C.
  {Helly}, A.~{Jenkins}, Y.~M. {Rosas-Guevara}, S.~D.~M. {White}, and J.~W.
  {Trayford}, {\it {The EAGLE simulations of galaxy formation: calibration of
  subgrid physics and model variations}},  {\em \mnras} {\bf 450} (Jun, 2015)
  1937--1961, [\href{http://arxiv.org/abs/1501.01311}{{\tt arXiv:1501.01311}}].

\bibitem{Heinrich:2016ojb}
C.~H. Heinrich, V.~Miranda, and W.~Hu, {\it {Complete Reionization Constraints
  from Planck 2015 Polarization}},  {\em Phys. Rev.} {\bf D95} (2017), no.~2
  023513, [\href{http://arxiv.org/abs/1609.04788}{{\tt arXiv:1609.04788}}].

\bibitem{Aghanim:2018eyx}
{\bf Planck} Collaboration, N.~Aghanim et~al., {\it {Planck 2018 results. VI.
  Cosmological parameters}},  [\href{http://arxiv.org/abs/1807.06209}{{\tt
  arXiv:1807.06209}}].

\bibitem{Zenodo}
A.~A. García, K.~Bondarenko, S.~Ploeckinger, J.~Pradler, and A.~Sokolenko,
  {\it {Electron number density and DM density along random lines of sight from
  the EAGLE simulation}},  Mar., 2020.
\newblock {Data is available at this URL:
  \url{https://zenodo.org/record/3715028} at Zenodo platform}.

\bibitem{Okun:1982xi}
L.~B. Okun, {\it {Limits of electrodynamics: paraphotons?}},  {\em Sov. Phys.
  JETP} {\bf 56} (1982) 502. [Zh. Eksp. Teor. Fiz.83,892(1982)].

\bibitem{Galison:1983pa}
P.~Galison and A.~Manohar, {\it {Two Z's or not two Z's?}},  {\em Phys. Lett.}
  {\bf 136B} (1984) 279--283.

\bibitem{Holdom:1985ag}
B.~Holdom, {\it {Two U(1)'s and Epsilon Charge Shifts}},  {\em Phys. Lett.}
  {\bf 166B} (1986) 196--198.

\bibitem{Georgi:1983sy}
H.~Georgi, P.~H. Ginsparg, and S.~L. Glashow, {\it {Photon Oscillations and the
  Cosmic Background Radiation}},  {\em Nature} {\bf 306} (1983) 765--766.

\bibitem{Jaeckel:2008fi}
J.~Jaeckel, J.~Redondo, and A.~Ringwald, {\it {Signatures of a hidden cosmic
  microwave background}},  {\em Phys. Rev. Lett.} {\bf 101} (2008) 131801,
  [\href{http://arxiv.org/abs/0804.4157}{{\tt arXiv:0804.4157}}].

\bibitem{Mirizzi:2009iz}
A.~Mirizzi, J.~Redondo, and G.~Sigl, {\it {Microwave Background Constraints on
  Mixing of Photons with Hidden Photons}},  {\em JCAP} {\bf 0903} (2009) 026,
  [\href{http://arxiv.org/abs/0901.0014}{{\tt arXiv:0901.0014}}].

\bibitem{Kunze:2015noa}
K.~E. Kunze and M.~A. Vazquez-Mozo, {\it {Constraints on hidden photons from
  current and future observations of CMB spectral distortions}},  {\em JCAP}
  {\bf 1512} (2015), no.~12 028, [\href{http://arxiv.org/abs/1507.02614}{{\tt
  arXiv:1507.02614}}].

\bibitem{McDermott:2019lch}
S.~D. McDermott and S.~J. Witte, {\it {The Cosmological Evolution of Light Dark
  Photon Dark Matter}},  [\href{http://arxiv.org/abs/1911.05086}{{\tt
  arXiv:1911.05086}}].

\bibitem{Nelson:2011sf}
A.~E. Nelson and J.~Scholtz, {\it {Dark Light, Dark Matter and the Misalignment
  Mechanism}},  {\em Phys. Rev.} {\bf D84} (2011) 103501,
  [\href{http://arxiv.org/abs/1105.2812}{{\tt arXiv:1105.2812}}].

\bibitem{Arias:2012az}
P.~Arias, D.~Cadamuro, M.~Goodsell, J.~Jaeckel, J.~Redondo, and A.~Ringwald,
  {\it {WISPy Cold Dark Matter}},  {\em JCAP} {\bf 1206} (2012) 013,
  [\href{http://arxiv.org/abs/1201.5902}{{\tt arXiv:1201.5902}}].

\bibitem{Dubovsky:2015cca}
S.~Dubovsky and G.~Hernández-Chifflet, {\it {Heating up the Galaxy with Hidden
  Photons}},  {\em JCAP} {\bf 1512} (2015), no.~12 054,
  [\href{http://arxiv.org/abs/1509.00039}{{\tt arXiv:1509.00039}}].

\bibitem{Graham:2015rva}
P.~W. Graham, J.~Mardon, and S.~Rajendran, {\it {Vector Dark Matter from
  Inflationary Fluctuations}},  {\em Phys. Rev.} {\bf D93} (2016), no.~10
  103520, [\href{http://arxiv.org/abs/1504.02102}{{\tt arXiv:1504.02102}}].

\bibitem{Kovetz:2018zes}
E.~D. Kovetz, I.~Cholis, and D.~E. Kaplan, {\it {Bounds on ultralight
  hidden-photon dark matter from observation of the 21 cm signal at cosmic
  dawn}},  {\em Phys. Rev.} {\bf D99} (2019), no.~12 123511,
  [\href{http://arxiv.org/abs/1809.01139}{{\tt arXiv:1809.01139}}].

\bibitem{Agrawal:2018vin}
P.~Agrawal, N.~Kitajima, M.~Reece, T.~Sekiguchi, and F.~Takahashi, {\it {Relic
  Abundance of Dark Photon Dark Matter}},  {\em Phys. Lett.} {\bf B801} (2020)
  135136, [\href{http://arxiv.org/abs/1810.07188}{{\tt arXiv:1810.07188}}].

\bibitem{Wadekar:2019xnf}
D.~Wadekar and G.~R. Farrar, {\it {First direct astrophysical constraints on
  dark matter interactions with ordinary matter at very low velocities}},
  [\href{http://arxiv.org/abs/1903.12190}{{\tt arXiv:1903.12190}}].

\bibitem{AlonsoAlvarez:2019cgw}
G.~Alonso-Álvarez, J.~Jaeckel, and T.~Hugle, {\it {Misalignment \& Co.:
  (Pseudo-)scalar and vector dark matter with curvature couplings}},  {\em
  JCAP} {\bf 2002} (2020), no.~02 014,
  [\href{http://arxiv.org/abs/1905.09836}{{\tt arXiv:1905.09836}}].

\bibitem{Caputo:2020bdy}
A.~Caputo, H.~Liu, S.~Mishra-Sharma, and J.~T. Ruderman, {\it {Dark Photon
  Oscillations in Our Inhomogeneous Universe}},
  [\href{http://arxiv.org/abs/2002.05165}{{\tt arXiv:2002.05165}}].

\bibitem{Bondarenko:2020moh}
K.~Bondarenko, J.~Pradler, and A.~Sokolenko, {\it {Constraining dark photons
  and their connection to 21 cm cosmology with CMB data}},
  [\href{http://arxiv.org/abs/2002.08942}{{\tt arXiv:2002.08942}}].

\bibitem{Pospelov:2018kdh}
M.~Pospelov, J.~Pradler, J.~T. Ruderman, and A.~Urbano, {\it {Room for New
  Physics in the Rayleigh-Jeans Tail of the Cosmic Microwave Background}},
  {\em Phys. Rev. Lett.} {\bf 121} (2018), no.~3 031103,
  [\href{http://arxiv.org/abs/1803.07048}{{\tt arXiv:1803.07048}}].

\bibitem{Bowman:2018yin}
J.~D. Bowman, A.~E.~E. Rogers, R.~A. Monsalve, T.~J. Mozdzen, and N.~Mahesh,
  {\it {An absorption profile centred at 78 megahertz in the sky-averaged
  spectrum}},  {\em Nature} {\bf 555} (2018), no.~7694 67--70,
  [\href{http://arxiv.org/abs/1810.05912}{{\tt arXiv:1810.05912}}].

\bibitem{planck2013}
{\bf Planck} Collaboration, Y.~Akrami et~al., {\it {Planck 2013 results. XVI.
  Cosmological parameters}},  {\em \aap} {\bf 571} (Nov., 2014) A16,
  [\href{http://arxiv.org/abs/1303.5076}{{\tt arXiv:1303.5076}}].

\bibitem{EagleDR}
{The EAGLE team}, {\it {The EAGLE simulations of galaxy formation: Public
  release of particle data}},  {\em arXiv e-prints} (Jun, 2017)
  arXiv:1706.09899, [\href{http://arxiv.org/abs/1706.09899}{{\tt
  arXiv:1706.09899}}].

\bibitem{2015MNRAS.447..499M}
I.~D. {McGreer}, A.~{Mesinger}, and V.~{D'Odorico}, {\it {Model-independent
  evidence in favour of an end to reionization by z {\ensuremath{\approx}} 6}},
   {\em \mnras} {\bf 447} (Feb., 2015) 499--505,
  [\href{http://arxiv.org/abs/1411.5375}{{\tt arXiv:1411.5375}}].

\bibitem{Fukugita:2004ee}
M.~Fukugita and P.~J.~E. Peebles, {\it {The Cosmic energy inventory}},  {\em
  Astrophys. J.} {\bf 616} (2004) 643--668,
  [\href{http://arxiv.org/abs/astro-ph/0406095}{{\tt astro-ph/0406095}}].

\bibitem{1999ApJ...523L...1S}
S.~{Seager}, D.~D. {Sasselov}, and D.~{Scott}, {\it {A New Calculation of the
  Recombination Epoch}},  {\em \apjl} {\bf 523} (Sep, 1999) L1--L5,
  [\href{http://arxiv.org/abs/astro-ph/9909275}{{\tt astro-ph/9909275}}].

\bibitem{Seager:1999km}
S.~Seager, D.~D. Sasselov, and D.~Scott, {\it {How exactly did the universe
  become neutral?}},  {\em Astrophys. J. Suppl.} {\bf 128} (2000) 407--430,
  [\href{http://arxiv.org/abs/astro-ph/9912182}{{\tt astro-ph/9912182}}].

\bibitem{pysph}
A.~Benitez-Llambay, {\it py-sphviewer: Py-sphviewer v1.0.0},  July, 2015.

\bibitem{cloudy17}
G.~J. {Ferland}, M.~{Chatzikos}, F.~{Guzm{\'a}n}, M.~L. {Lykins}, P.~A.~M. {van
  Hoof}, R.~J.~R. {Williams}, N.~P. {Abel}, N.~R. {Badnell}, F.~P. {Keenan},
  R.~L. {Porter}, and P.~C. {Stancil}, {\it {The 2017 Release Cloudy}},  {\em
  \rmxaa} {\bf 53} (Oct, 2017) 385--438,
  [\href{http://arxiv.org/abs/1705.10877}{{\tt arXiv:1705.10877}}].

\bibitem{hm12}
F.~{Haardt} and P.~{Madau}, {\it {Radiative Transfer in a Clumpy Universe. IV.
  New Synthesis Models of the Cosmic UV/X-Ray Background}},  {\em \apj} {\bf
  746} (Feb, 2012) 125, [\href{http://arxiv.org/abs/1105.2039}{{\tt
  arXiv:1105.2039}}].

\bibitem{rahmati2013}
A.~{Rahmati}, A.~H. {Pawlik}, M.~{Rai{\v{c}}evi{\'c}}, and J.~{Schaye}, {\it
  {On the evolution of the H I column density distribution in cosmological
  simulations}},  {\em \mnras} {\bf 430} (Apr., 2013) 2427--2445,
  [\href{http://arxiv.org/abs/1210.7808}{{\tt arXiv:1210.7808}}].

\bibitem{Wolfire:1995fe}
M.~G. Wolfire, D.~Hollenbach, C.~F. McKee, A.~G. G.~M. Tielens, and E.~L.~O.
  Bakes, {\it {The neutral atomic phases of the interstellar medium}},  {\em
  Astrophys. J.} {\bf 443} (1995) 152--168.

\bibitem{Wolfire:2002jm}
M.~G. Wolfire, C.~F. McKee, D.~Hollenbach, and A.~G. G.~M. Tielens, {\it
  {Neutral atomic phases of the ISM in the galaxy}},  {\em Astrophys. J.} {\bf
  587} (2003) 278--311, [\href{http://arxiv.org/abs/astro-ph/0207098}{{\tt
  astro-ph/0207098}}].

\bibitem{An:2013yfc}
H.~An, M.~Pospelov, and J.~Pradler, {\it {New stellar constraints on dark
  photons}},  {\em Phys. Lett.} {\bf B725} (2013) 190--195,
  [\href{http://arxiv.org/abs/1302.3884}{{\tt arXiv:1302.3884}}].

\bibitem{Fixsen:1996nj}
D.~J. Fixsen, E.~S. Cheng, J.~M. Gales, J.~C. Mather, R.~A. Shafer, and E.~L.
  Wright, {\it {The Cosmic Microwave Background spectrum from the full COBE
  FIRAS data set}},  {\em Astrophys. J.} {\bf 473} (1996) 576,
  [\href{http://arxiv.org/abs/astro-ph/9605054}{{\tt astro-ph/9605054}}].

\bibitem{2018arXiv180706206P}
{\bf Planck} Collaboration, Y.~Akrami et~al., {\it {Planck 2018 results. II.
  Low Frequency Instrument data processing}},  {\em arXiv e-prints} (Jul, 2018)
  arXiv:1807.06206, [\href{http://arxiv.org/abs/1807.06206}{{\tt
  arXiv:1807.06206}}].

\bibitem{article_SPT}
E.~George, C.~Reichardt, K.~Aird, B.~Benson, L.~Bleem, J.~Carlstrom, C.~Chang,
  H.-M. Cho, T.~Crawford, A.~Crites, T.~Haan, M.~Dobbs, J.~Dudley,
  N.~Halverson, N.~Harrington, G.~Holder, W.~Holzapfel, Z.~Hou, J.~Hrubes, and
  O.~Zahn, {\it A measurement of secondary cosmic microwave background
  anisotropies from the 2500 square-degree spt-sz survey},  {\em The
  Astrophysical Journal} {\bf 799} (08, 2014).

\bibitem{Poulin:2016nat}
V.~Poulin, P.~D. Serpico, and J.~Lesgourgues, {\it {A fresh look at linear
  cosmological constraints on a decaying dark matter component}},  {\em JCAP}
  {\bf 1608} (2016), no.~08 036, [\href{http://arxiv.org/abs/1606.02073}{{\tt
  arXiv:1606.02073}}].

\bibitem{Chluba:2015hma}
J.~Chluba, {\it {Green's function of the cosmological thermalization problem --
  II. Effect of photon injection and constraints}},  {\em Mon. Not. Roy.
  Astron. Soc.} {\bf 454} (2015), no.~4 4182--4196,
  [\href{http://arxiv.org/abs/1506.06582}{{\tt arXiv:1506.06582}}].

\bibitem{Cui:2017ytb}
Y.~Cui, M.~Pospelov, and J.~Pradler, {\it {Signatures of Dark Radiation in
  Neutrino and Dark Matter Detectors}},  {\em Phys. Rev.} {\bf D97} (2018),
  no.~10 103004, [\href{http://arxiv.org/abs/1711.04531}{{\tt
  arXiv:1711.04531}}].

\bibitem{2017MNRAS.466..960P}
A.~H. {Pawlik}, A.~{Rahmati}, J.~{Schaye}, M.~{Jeon}, and C.~{Dalla Vecchia},
  {\it {The Aurora radiation-hydrodynamical simulations of reionization:
  calibration and first results}},  {\em \mnras} {\bf 466} (Apr, 2017)
  960--973, [\href{http://arxiv.org/abs/1603.00034}{{\tt arXiv:1603.00034}}].

\bibitem{2013A&A...556A...2V}
M.~P. {van Haarlem} et~al., {\it {LOFAR: The LOw-Frequency ARray}},  {\em \aap}
  {\bf 556} (Aug., 2013) A2, [\href{http://arxiv.org/abs/1305.3550}{{\tt
  arXiv:1305.3550}}].

\bibitem{Bacon:2018dui}
{\bf SKA} Collaboration, D.~J. Bacon et~al., {\it {Cosmology with Phase 1 of
  the Square Kilometre Array: Red Book 2018: Technical specifications and
  performance forecasts}},  {\em Submitted to: Publ. Astron. Soc. Austral.}
  (2018) [\href{http://arxiv.org/abs/1811.02743}{{\tt arXiv:1811.02743}}].

\bibitem{Braaten:1993jw}
E.~Braaten and D.~Segel, {\it {Neutrino energy loss from the plasma process at
  all temperatures and densities}},  {\em Phys. Rev.} {\bf D48} (1993)
  1478--1491, [\href{http://arxiv.org/abs/hep-ph/9302213}{{\tt
  hep-ph/9302213}}].

\bibitem{Peck:77}
E.~R. Peck and S.~Huang, {\it Refractivity and dispersion of hydrogen in the
  visible and near infrared},  {\em J. Opt. Soc. Am.} {\bf 67} (Nov, 1977)
  1550--1554.

\end{thebibliography}\endgroup

\end{document}